\documentclass[%
	a4paper,
	twoside,
	bibliography=totoc,
	listof=totoc,
	index=totoc,
	parskip=half,
	chapterprefix=false,
	numbers=noenddot,
	11pt
]{article} 

\usepackage{authblk}
\usepackage[left=3cm,right=2.5cm,top=3cm,bottom=4cm]{geometry}  
\usepackage{rotating}                                           
\usepackage[singlespacing]{setspace}                           
\usepackage{indentfirst}                                        
\usepackage{titlesec}                                           
\titlespacing{\section}{0pt}{36pt}{6pt}                         
\titlespacing{\subsection}{0pt}{24pt}{6pt}                      
\usepackage[toc,page]{appendix}                                 
\usepackage{parskip}

\usepackage{color}						

\usepackage{scrpage2}					        
\clearscrheadfoot                                               
\setheadsepline{1pt}                                            
\setfootsepline{1pt}                                            
\lohead[]{\leftmark}                                            
\rehead[]{\rightmark}                                           

\usepackage{array}						
\usepackage{booktabs}                   			
\usepackage{longtable}						
\usepackage{rotating}                                           
\usepackage{multirow}						
\usepackage{todonotes}

\usepackage{graphicx}						
\usepackage{subcaption}                     			
\usepackage{caption}						
\graphicspath{{images/}{plots/}{figures/}}

\usepackage{amsfonts}						
\usepackage{amssymb}						
\usepackage{amsmath}						
\usepackage{gensymb}

\usepackage[authoryear]{natbib}                                 
\usepackage{acronym}                                            
\usepackage{url}

\newcommand{\unit}[1]{\ensuremath{\, \mathrm{#1}}}

\begin{document}

\newcommand{\HRule}{\rule{\linewidth}{0.5mm}}

\title{A Large Deviation Theory-based Analysis of Heat Waves and Cold Spells in a Simplified Model of the General Circulation of the Atmosphere}
\author{Vera Melinda G\'alfi (1,2), Valerio Lucarini (1,3,4), Jeroen Wouters (3,5)}

\affil{(1) Meteorological Institute, CEN, University of Hamburg, Hamburg, Germany, (2) IMPRS-ESM, Max Planck Institute for Meteorology, Hamburg, Germany, (3) Department of Mathematics and Statistics, University of Reading, Reading, UK, (4) Center for the Mathematics of Planet Earth, University of Reading, Reading, UK,  (5) Niels Bohr Institute, University of Copenhagen, Copenhagen, Denmark}

\maketitle

\pagenumbering{arabic}
\setcounter{page}{1}
\cfoot[\pagemark]{\pagemark}                                    


\begin{abstract}

We study temporally persistent and spatially extended extreme events of temperature anomalies, i.e. heat waves and cold spells, using large deviation theory. To this end, we consider a simplified yet Earth-like general circulation model of the atmosphere and numerically estimate large deviation rate functions of near-surface temperature in the mid-latitudes.  We find that, after a re-normalisation based on the integrated auto-correlation, the rate function one obtains at a given latitude by looking, locally in space, at long time averages  agrees with what is obtained, instead, by looking, locally in time, at large spatial averages along the latitude. This is a result of scale symmetry in the spatial-temporal turbulence and of the fact that advection is primarily zonal. This agreement hints at the universality of large deviations of the temperature field. Furthermore, we discover that the obtained rate function is able to describe spatially extended and temporally persistent heat waves or cold spells, if we consider temporal averages of spatial averages over intermediate spatial scales. Finally, we find out that large deviations are relatively more likely to occur when looking at these spatial averages performed over intermediate scales, thus pointing to the existence of weather patterns associated to the low-frequency variability of the atmosphere. Extreme value theory is used to benchmark our results. 


\end{abstract}
\section{Introduction and Motivation}
\label{sec:intro}

The typical way to formalise the analysis of extremes for a stochastic variable $X$ revolves around looking at the tail of the probability distribution of $X$ and identifying extremes as very large (or very small) events with long return time. This point - as discussed below - is mathematically very powerful, but in the usual setting is not well suited for studying, in the case of spatio-temporal chaos, anomalously large or small events that are persistent in time and extended in space. 
\textit{Persistent} climatic \textit{extreme events} - like heat waves or cold spells - can have a huge impact: they do not affect only human health, but also ecosystems; they can be a danger for our infrastructures and crops, and have a destabilising effect over entire societies; the scale of damage depends critically on the persistent nature and spatial extent of the events.  \citep{easterlingetal2000,robinson2001,who2004,ipcc2012}. Among the most relevant historical examples we would like to mention the mega-drought that played a major role in the collapse of the Maya empire \citep{Kennett2012}, and the recurrent extreme cold spell episodes referred to as Dzud that led to various waves of migration of the nomadic Mongolian populations \citep{Fang1992, Hvistendahl2012}.

\begin{figure}[ht!]
\begin{subfigure}{0.5\textwidth}
	a)\\
	\includegraphics[width=0.9\linewidth, height=5cm]{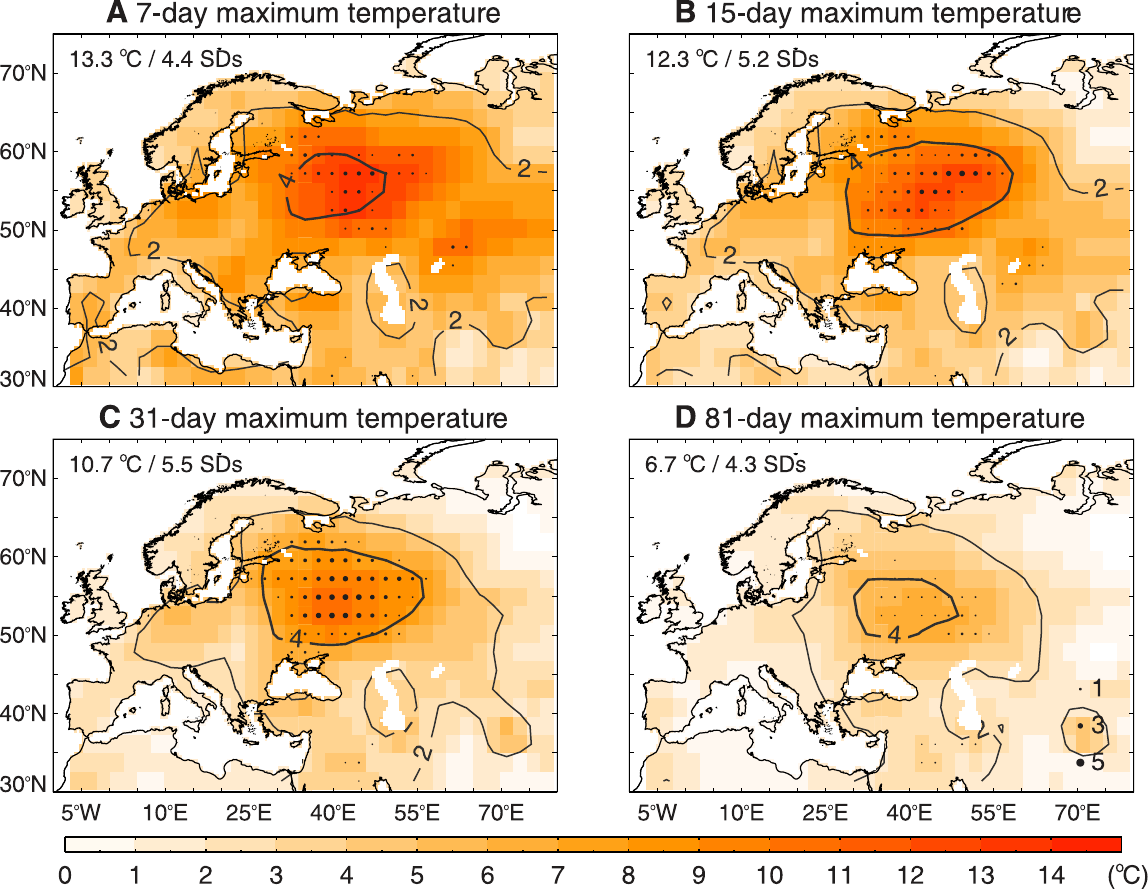} 
	\label{fig:subim1}
\end{subfigure}
\begin{subfigure}{0.5\textwidth}
	b)\\
	\includegraphics[width=0.9\linewidth, height=5cm]{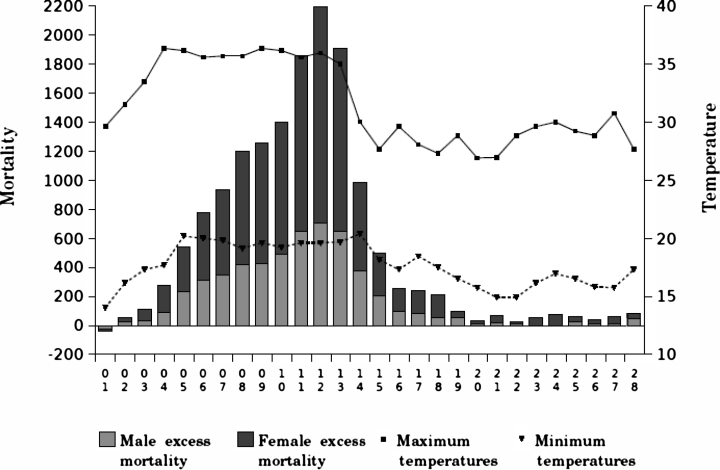}
	\label{fig:subim2}
\end{subfigure}
\caption{Spatio-temporal features of heat waves and their impacts on health. a) Anomalies of temperature maxima over four different time scales during the 2010 heat wave. Contour lines indicate the anomalies divided by the climatological standard deviation of temperature in the same location during summer days. The record breaking locations are indicated with dots. The absolute maxima are indicated in the top left corners. Reproduced from \citet{barriopedro2011}. b) Number of excess deaths in France in August (calendar days of the month indicated in the x-axis) during the 2003 heat wave. Note that the number of daily excess deaths increases day by day during the heat wave, and goes rapidly to zero afterwards. Reproduced from \cite{Poumadere2005}.} 
  \label{fig;heatwave}
\end{figure}

A heat wave or a cold spell is not only lasting for a long time (from several days to several weeks, even months) but has also a considerable spatial extent. For example, the 2003 and 2010 European heat waves had a temporal and spatial extent of the order of weeks to months and $10^6 \unit{km^2}$ respectively \citep{barriopedro2011}. These persistent events are primarily caused by anomalous synoptic conditions, and, in the case of the mid-latitudes, by atmospheric blocking situations, so we talk about persistence in space and time on \textit{large synoptic scales} \citep{vautardetal2007,sillmannetal2011,stefanonetal2012}. In Fig.~\ref{fig;heatwave}a we portray the intensity and extent of the 2010 heat wave and, in Fig.~\ref{fig;heatwave}b, we show how the excess fatalities observed in France during the 2003 heat wave increase dramatically as a result of persistent large positive anomalies of temperatures  \citep{Poumadere2005}.

In the climate system, there is a non-trivial relationship between spatial and temporal scales of variability - with large spatial scales typically associated to longer time scales. An effective way to represent such relationship is through the so-called Hayashi spectra \citep{Hayashi1971,Fraedrich1978,Speranza1983,DellAquila2005}. The existence of such relationship comes from the fact that one can loosely identify different dynamical regimes, each characterised by specialised dynamical balances between the forces acting on the fluid components \citep{lucarinietal2014}. Such balances can be rigorously derived via asymptotic analysis applied on the Navier-Stokes equations on a rotating frame of reference \citep{klein2010}. 

As well known, the understanding of the processes associated to synoptic disturbances, which dominate the weather variability at mid-latitudes is firmly grounded in the theory of baroclinic instability (see \citealt{holton2004}). Things are considerably less clear when looking into phenomena characterised by longer characteristic time scales.

Often the term low-frequency variability is used to describe a vast range of atmospheric processes occurring on a time scale ranging from about a week to about a month. Low-frequency variability features a much greater variety of phenomena with respect to synoptic variability, and, despite decades of efforts in terms of theoretical studies, observations, and numerical modelling, no complete understanding has yet been reached. Persistent weather anomalies, which can lead to long-lasting temperature extremes, i.e. heat waves and cold spells,  are associated to quasi-stationary Rossby waves \citep{sillmannetal2011,stefanonetal2012}.
The phase speed of these waves depends on the wavelength and is always westwards, i.e. opposite to the direction of the mean flow, at mid-latitudes. If the wavelength is large enough, the phase speed of Rossby waves can become very low or even zero, giving rise to quasi-stationary or stationary anomalous synoptic situations, so-called blocking events (see the recent review by \citealp{tibaldimolteni2018}).

\subsection{A Mathematical Framework for Climatic Extremes}
\subsubsection{Extreme Value Theory}
A robust mathematical framework for the study of extreme values is provided by extreme value theory. One way to construct a theory of extremes according to the procedure of block maxima can be summarised as follows. One considers a sequence of realisations of independent and identically distributed variables $X_1, X_2,...$ and takes the maximum $M_m$ over $m$ of such variables \citep{fisher1928,gnedenko1943}. Alternatively, extremes can be constructed according to the so-called peak over threshold method by considering the same sequence of variables as before, and selecting the values exceeding a given threshold $u$ \citep{pickands1975,balkemadehaan1974}. Both methods are formulated in form of limit laws, and rely on the convergence in distribution of the selected extreme values to one limiting family of distributions for a vast class of parent distributions, as one considers more and more extreme levels, i.e. for the block sizes $m$ and threshold $u$, respectively. The condition of independence of the stochastic variables can be relaxed in order to include the case of weakly correlated variables \citep{leadbetteretal1983}, and can be formulated in such a way to allow the establishment of extreme value laws for observables of chaotic dynamical systems \citep{lucarini2016}. 

The limiting family of distributions is the Generalised Extreme Value distribution in case of the block maxima method, and the Generalised Pareto distribution in case of the peak over threshold approach. The Pickands-Balkema-de Haan theorem guarantees that, in the limit, there is a one-to-one correspondence between the Generalised Extreme Value and Generalised Pareto limiting distributions for a given dataset, even if, when finite data are considered, the two approaches select different candidates for extremes \citep{pickands1975,balkemadehaan1974}. If the limiting Generalised Extreme Value or Generalised Pareto distributions exist and can be estimated reliably, one can calculate the probability of occurrence of events that are \textit{more extreme} than any observed event. In other words, reliable estimates of return periods for time ranges longer than what is actually observed can be constructed. This indicates that if the limit law applies, predictive power (in a statistical sense) emerges for events with very low probability of occurrence. Note that while universality emerges in the limit, the speed at which the asymptotic properties are realised are process-specific, i.e. not universal \citep{galfietal2017}. 

As clearly emerges from the discussion above, a statistical analysis based on Extreme Value Theory is in principle extremely useful exactly for those stakeholders that need to plan for long time ahead, and has in fact long been applied in areas such as finance \citep{embrechts}, engineering \citep{castillo}, and hydrology \citep{Katz2002}. Somewhat surprisingly,  while examples of application can obviously be found, the methods of Extreme Value Theory are still not the mainstream approach for studying very intense events in climate studies, where it is more popular to use empirical methods based on the analysis of high (or low) percentiles of the  probability distribution of the variable of interest. In fact, it is usually assumed that the theory is too data-hungry for being effectively applied in most available climatic time series \citep{ipcc2012}. Nonetheless, it has been recently shown that the theory of extreme values can be rigorously and robustly applied also in case of relatively short time series of a few tens of years, see e.g. \citet{zahidetal2017}.

We have mentioned above the problem of persistence. One can analyse persistent events generally in two ways: first, by treating them as a concatenation of successive extreme events and study the properties of clusters of extremes \citep{ferrosegers2003,segers2005}, or, second, by looking at pdf's of \textit{time-averaged observables}. In this study we follow the second route. 
Following intuition, if we look at the pdf of finite-size averages of an observable, one expects that the tails of the distribution are mainly populated by averages coming from persistent extremes. A rationale for this is that the averaging window acts like a low-pass filter on the length of the considered persistent events, leading to a connection between extremes of averages and persistent events with a certain length (greater or approximately equal to the chosen averaging window). This will roughly be, in fact, the scenario we will explore below. However, the link between persistence and extremes of finite-size averages is not always true: in case of heavy-tailed random variables, for example, the extremes of averages  are dominated by a single very large extreme event within the averaging window \citep{mikoschnagaev1998}.
We remark that, generally, the methods of Extreme Value Theory can be applied the same way also to study extremes of averaged observables. However, the averaging process reduces the number of available data, so that these methods can become more difficult to apply as the averaging window length is increased.

\subsubsection{Large Deviation Theory}
 A mathematical framework describing properties of pdf's of \textit{averaged} observables is provided by large deviation theory (LDT), introduced by \citet{cramer1938} and further developed by other mathematicians, like \citet{donskervaradhan1975_1,donskervaradhan1975_2,donskervaradhan1976,donskervaradhan1983}, \citet{gaertner1977}, and \citet{ellis1984}. The central result of the theory consists of writing the probability of averaged random variables $A_n=\frac{1}{n}\sum_{i=1}^{n} X_i$ as function of the variables $X_i$: for $n\to \infty$ the probability of averages decays exponentially with $n$, $p(A_n=a)\approx \mathrm{e}^{-n I(a)}$. This is called a \textit{large deviation principle}. The speed of decay is described by the so called rate function $I(a)\ge 0$. The probability $p(A_n=a)$ decays everywhere with increasing $n$, except when $I(a)=0$. Here $\mathrm{e}^{-n I(a)}=1$. For independent identically distributed random variables, one would have that $\mathbb{E}[A_n]=a^*$, where $a^*$ is such that $I(a^*)=0$. If the rate function exists, one can estimate the probability of averages for every $n$. Similarly to extreme value theory, if the limit law applies, we gain predictive power, with the difference that in this case it is directed towards \textit{averages with increasing} $n$. This means that one does not have to deal anymore with the problem of decreasing amount of data as $n$ increases. The theory of large deviations is used very extensively in physics, mostly in the context of thermodynamics and statistical mechanics.

While they have recently been applied in the context of geophysical flows (see e.g. \citealt{bouchetvenaille2012,bouchetetal2014,herbert2015}), techniques of LDT have been used sporadically until now in climate studies, despite the fact that they can be useful whenever the connection between macroscopic or long term observables and microscopic or instantaneous observables is important, and one is interested in persistent and/or extended fluctuations of a climatic field. 

One area of climate modelling where techniques based on large deviations are just beginning to be applied is the sampling of rare events. Rare event computation techniques based on elements of LDT have been developed with the aim to produce reliable statistics of specific rare events of a given model, as an alternative to long direct numerical simulations \citep{giardinaetal2011, woutersbouchet2016,lestangetal2018}. \citet{ragoneetal2017} describes how model trajectories can be selected, based on a rare event algorithm, by keeping an ensemble realisation of the system in states that are preferentially close to those leading to heat waves. Therefore, one can exponentially oversample events that have ultra long return periods, and thus construct a richer statistics of heat waves than one would get by standard Monte Carlo techniques. The described method provides also the possibility to investigate dynamical properties of the system state (like global circulation patterns and jet stream position) supporting the occurrence of the studied extremes (heat waves).

\subsection{This Paper}
In this work we adopt LDT to analyse the properties of temporally and/or spatially persistent surface temperature extremes - heat waves or cold spells - generated through simulation performed with  the Portable University Model of the Atmosphere (PUMA) \citep{lunkeitetal1998,fraedrichetal2005a}. We  investigate temperature averages computed in time and/or in space, the spatial averaging being performed along the zonal direction for reasons of symmetry. We point out that this is the first study to analyse persistent climatic events based on this simple application of LDT. We perform non-equilibrium steady state model simulations using idealised conditions, which cannot be directly related to realistic atmospheric states. Thus, the endeavour in this work is to test the introduced methodology, and to understand the possible potential by applying these methods of statistical mechanics to the atmosphere, rather than analysing real atmospheric conditions. However, this work should be also seen as a first step into the direction of applying LDT to more realistic climate simulations and to observational data. At this stage, we do not investigate the dynamical processes leading to the heat waves and cold spells, but rather try to construct their asymptotic statistical properties. 

PUMA - details given in Sec.~\ref{sec:model} - describes with a good level of precision the dynamics of the three-dimensional atmosphere as an out-of-equilibrium forced-dissipative system. We analyse the properties of the steady state achieved as a result of time-independent forcing after transient dynamics has been discarded. For a wide range of  parameter values PUMA features  high-dimensional chaotic dynamics \citep{decruzetal2018}. By considering the connection between the averaged values and persistent events on suitably defined scales (as explained above), large deviations of temperature can possibly be related to persistent extreme events of temperature, i.e. heat waves or cold spells. 

Following the discussion above on the phenomenology of synoptic disturbances, we expect to find a link between spatially extended and temporally persistent events. In order to achieve a large deviation when considering spatial averages in a turbulent system, we need to have occurrence of a spatially extended structure of length say $L$. In a system possessing a characteristic velocity scale $U$ one expects such a structure to persist for a typical time of the order $\tau=L/U$. In this work, we explore the connection between temporal and spatial large deviations, and we also analyse spatio-temporal large deviations. We seek answers to two main questions:
\begin{enumerate}
\item How well does LDT describe persistent in space and/or time temperature fluctuations in PUMA?
\item What is the link between temporal, spatial, and spatio-temporal large deviations?
\end{enumerate}
These questions are potentially relevant, because, if we find experimental proofs that the large deviation limit does hold in the case of our numerical simulations, there is a good chance to calculate the probability of occurrence of arbitrarily long in time and/or extended in space (within the limits allowed by the geometry of the Earth, as seen later) heat waves and cold spells. We point out that the possibility of establishing large deviation laws in geophysical systems is a non-trivial matter, as a result of the presence of temporal and spatial correlations on multiple scales. The strength of these correlations is crucial for the practical applicability of the theory given a finite amount of data. We remark that, when considering coupled atmospheric and oceanic dynamics, finding large deviation laws can become a difficult task. Examining dynamical indicators, \citet{decruzetal2018} could not detect large deviations laws in case of finite time Lyapunov exponents in a quasi-geostrophic coupled ocean-atmosphere model. \citet{vannitsemlucarini2016} analysed the large deviations of finite time Lyapunov exponents as well in a low-order version of the above mentioned coupled model, and found a large deviation principle only in case of nonzero Lyapunov exponents, whereas the convergence was considerably slower or even absent in case of near-zero Lyapunov exponents.

Our model does not feature the presence of slow oceanic time scales, and, therefore, provides a simpler setting to test our ideas. In the case we find a link between temporal and spatial large deviations, we can deduce the probability of spatial (or spatio-temporal) averages from the one of temporal averages and vice-versa. This can be very useful in case of applications, when for example only temporal or only spatial series are available. In order to test the quality of the predictions of the return time based on LDT, we compare the results with what can be obtained using extreme value theory (we use the peak over threshold method). 

The structure of the paper is as follows. In Section \ref{sec:theory}, we provide the theoretical formulation of LDT and some elements of extreme value theory. In Section \ref{sec:model}  we give a description of the model PUMA and give details of the numerical simulations performed for the scopes of this paper. We present our results in Section \ref{sec:results}. Here, we first focus on the link between temporal and spatial large deviations, and then we additionally consider the case of spatio-temporal large deviations. We test the correctness and applicability of our results by computing return periods of extremes of temperature averages and comparing them with the empirical data and with return periods obtained based on the peak over threshold method. Additionally, in order to assess the robustness and applicability of the proposed approach, we test how our findings related to return periods change when considering shorter time series for estimating the large deviation rate functions. Section \ref{sec:disc} concludes the paper containing a summary and discussion of our results and ideas for future investigations.
\section{Some Elements of LDT and of Extreme Value Theory}
\label{sec:theory}

\subsection{Constructing the Rate Functions Describing the Large Deviations}
The large deviation theoretical framework can be formulated on three different levels, corresponding to the complexity of the statistical description of the dynamical system. These are, as described by \citet{oono1989}, based on: sample means of observables (level-1), probability distributions on the state space of observables (level-2), and probability distributions on the path or history space, i.e. the entire set of possible orbits or histories of the system (level-3). The below description follows the level-1 approach, according to the scientific purpose of this paper, and is mostly based on the works of \citet{touchette2009} and \citet{oono1989}. We do not pursue at all a rigorous mathematical formulation here; our aim is rather to recapitulate the main concepts and results, and to introduce our notation.

We say that the random variable $A_n=\frac{1}{n}\sum_{i=1}^n X_i$, where $X_i$ are identically distributed random variables, satisfies a large deviation principle if the limit 
  \begin{equation}
   \lim_{n\to\infty}-\frac{1}{n}\ln p(A_n=a)=I(a)
   \label{eq:ldlim}
  \end{equation}
exists. The probability density $p(A_n=a)$ decays exponentially with $n$ for every value of $a$ except the ones for which $I(a)=0$, where $\lim_{n\to \infty}p(A_n=a^*)=1$, and $a^*=\mathbb{E}[A_n]$. $I(a)\ge 0$ is the so-called rate function, representing the rate of this exponential decay of probabilities. Whenever limit (\ref{eq:ldlim}) holds and $I(a)$ has a unique global minimum, $A_n$ converges in probability to its mean $a^*$ and obeys the law of large numbers. If then additionally $I(a)$ is quadratic (i.e. twice differentiable) around $a^*$, the central limit theorem holds, meaning that small fluctuations around the mean are normally distributed. The expression ``small fluctuations'' is very important here, because large fluctuations around the mean are not necessarily normally distributed. Since the rate function describes both small and large deviations, LDT can be considered as a generalisation of the central limit theorem.
  
Now let's consider, instead of random variables, observables produced by a deterministic dynamical system. If the system is Axiom A, all of its observables obey a large deviation principle \citep{eckmann85}. If we consider a high-dimensional chaotic system, by invoking the chaotic hypothesis introduced by \citet{gallavotticohen1995}, one can expect to find large deviation laws, even in systems which are not Axiom A.

The dynamical nature of out-of-equilibrium steady state systems requires, however, a slight modification of our theoretical approach, which mainly implies that time has to be considered in the formulation of the large deviation principle, replacing the parameter $n$. Due to temporal correlations in these systems the computation of the rate function requires level-2 or level-3 theory. This has been done for Markov chains and random variables with a specific form of dependence, and involves mostly the computation of transition matrices or joint pdf's \citep{denhollander2000,touchette2009}. In case of non-Markovian processes and high dimensional systems the computation of analytical rate functions is a hopeless endeavour. Thus, in this work, we adopt another (very simple) strategy in dealing with temporal correlations. In case of weakly correlated observables (i.e. $X_j$ and $X_l$ have an exponentially decreasing correlation if $|j-l|$ is large enough), one can take advantage of the fact that for large enough $n$ the averages $A_n$ become almost uncorrelated. This represents the basis for the block averaging method \citep{rohweretal2015}. We transform the variables $X_i$ into  variables $Y_i=\frac{1}{b}\sum_{i=1}^b X_i$, where $b$ represents the size of the averaging block, i.e $b=n/k$ with the number of blocks $k$. In case $Y_i$ are almost independent and identically distributed (ergodic Markov chain), a large deviation principle can be obtained for:
  \begin{equation}
   A_n=\frac{1}{k}\sum_{i=1}^k Y_i=\frac{1}{n}\sum_{i=1}^n X_i.
  \end{equation}
Intuitively, one can argue that $b$ has to be at least so large that $X_i$ and $X_{i+b}$ are almost uncorrelated, i.e. $b\ge \rho$ where $\rho$ is a metric of persistence expressed in terms of number of successive correlated data. One usually quantifies persistence in terms of the auto-correlation function. Considering our scientific goal, which is the study of probabilities of averages, it makes sense to choose the integrated auto-correlation as a general measure of persistence in time and space, since this quantity plays a central role in the central limit theorem for Markov chains, as described below.
      
According to a formulation of the central limit theorem in case of dependent variables based on \citet{billingsley1995}, suppose that $X_1,X_2,...$ is a stationary Markov chain with $\mathbb{E}[X_n]=0$ and satisfies appropriate mixing conditions, then the variance of the sample mean $A_n$ is
\begin{equation}
  \label{eq:cltmc}
  n \mathbb{E}[A_n^2] \to \mathbb{E}[X_1^2](1+2\sum_{k=1}^\infty c(k))
\end{equation}

where $c(k)=\frac{C(k)}{C(0)}$ is the auto-correlation, and $C(k)$ denotes the auto-covariance at lag $k$, $C(k)=\mathbb{E} [X_i X_{i+k}]$. 
Eq.~(\ref{eq:cltmc}) shows that the rescaled variance of the sample mean of the Markov chain converges to the variance of $X_1$ times the integrated auto-correlation.

It is well know that the central limit theorem is violated when large extreme values dominate the fluctuations around the mean. In these cases, the probability of sums converges to a more general limit instead of the Gaussian distribution. This limit is represented by the class of infinitely divisible distributions, including Levy alpha-stable distributions \citep{westetal2003}. As a consequence of diverging second (or even first) moments of the distribution of the stochastic variable of reference, the probability of sums decays sub-exponentially and the rate function is trivially 0 \citep{touchette2009}. Large deviations results can still be obtained in many cases, however they are dominated by the largest values in the sample instead of the mean, as already mentioned in the introduction \citep{mikoschnagaev1998}. These conditions are relevant for some variables of interest in (geo)physical fluid dynamics. It has been shown that for certain variables of turbulent flows the central limit theorem does not hold. For example, velocity differences (or gradients) between two points in space often develop long tails, as an effect of long-lived strong vortices near the dissipative range of scales \citep{biferale1993,jimenez1996,jimenez2000}. Several climatic variables have been also found to exhibit an increasing variability at low frequencies: atmospheric surface variables in the tropics (due to the effect of pulse-like convective events), or sea surface temperature in some regions \citep{fraedrichetal2009,blenderetal2008}. It is expected that in these cases the long term memory prevents convergence to what predicted by the central limit theorem, at least on the relevant finite scales.
  
\subsection{Extreme Value Theory: Peak over Threshold Approach}
  
A straightforward way to investigate the extremes of averaged quantities is, clearly, thorough the use of extreme value theory. Note that, despite in many practical applications such an approach is infeasible or not practical because the procedure of averaging reduces dramatically the size of the dataset, our numerical simulations are long enough to allow for a reliable implementation also in the case of averages. We briefly summarise below the main ideas of extreme value theory.  
  
Let us consider $Z_{m} = \mathrm{max}\{X_{1},...,X_{m}\}$, where $X_{1},...,X_{m}$ is a sequence of independent and identically distributed random variables with common distribution function $F(x)$. The Fisher-Tipett-Gnedenko theorem \citep{fisher1928,gnedenko1943} states that the distribution of properly normalised block maxima $Z_m$ converges under certain conditions, for $m \to \infty $, to the so-called Generalised Extreme Value distribution $G(z;\mu,\sigma,\xi)$, with three parameters: location parameter $\mu$, scale parameter $\sigma$, and shape parameter $\xi$:
\begin{equation}
  G(z) = 
  \begin{cases}
     \exp\left\{-\left[1+\xi\left(\frac{z-\mu}{\sigma}\right)\right]^{-1/\xi}\right\} & \quad \mathrm{for} \ \xi \ne 0,\\
     \exp\left\{-\exp\left[-\left(\frac{z-\mu}{\sigma}\right)\right]\right\}          & \quad \mathrm{for} \ \xi = 0,\\
  \end{cases}
  \label{eq:GEV1}
\end{equation} 
where $-\infty<\mu<\infty$, $\sigma>0$, $1+\xi(z-\mu)/\sigma>0$ for $\xi \ne 0$ and $-\infty<z<\infty$ for $\xi=0$ \citep{coles2001}.

\citet{pickands1975} reformulated the Fisher-Tipett-Gnedenko theorem based on the conditional probability of values exceeding a high threshold $u$ and reaching the upper right point of the distributions of $X$, given that $X>u$. Under the same conditions such that the distribution of $Z_{m}$ converges to the Generalised Extreme Value distribution, the exceedances $y=X-u$ are asymptotically distributed according to the Generalised Pareto distribution family \citep{coles2001}
\begin{equation}
  H(y; \tilde{\sigma},\tilde{\xi}) = 
  \begin{cases}
    1-\left(1+\frac{\tilde\xi y}{\tilde{\sigma}}\right)^{-1/\tilde\xi}     & \quad \mathrm{for} \ \tilde\xi \ne 0,\\
    1-\exp\left(-\frac{y}{\tilde{\sigma}}\right)               & \quad \mathrm{for} \ \tilde\xi = 0,\\
  \end{cases}
  \label{eq:GPD1}
\end{equation}
where $1+\tilde\xi y/\tilde{\sigma}>0$ for $\tilde\xi \ne 0$, $y>0$, and $\tilde{\sigma}>0$. $H(y)$ has two parameters: the scale parameter $\tilde{\sigma}$ and the shape parameter $\tilde\xi$. The shape parameter \(\tilde\xi\) describes the decay of probabilities in the tail of the distribution, and determines to which one of the three possible types of Generalised Pareto distributions $H(y)$ belongs. If $\tilde\xi=0$, the tail decay is exponential; if \(\tilde\xi>0\), the tail decay is polynomial; and if \(\tilde\xi<0\) the distribution is bounded, i.e. the extremes are limited from above \citep{pickands1975,balkemadehaan1974}. The parameters of the Generalised Extreme Value and Generalised Pareto distributions are related as follows: $\tilde\xi=\xi$ and $\tilde\sigma=\sigma+\xi(u-\mu)$ \citep{coles2001}. This implies that the two, block maxima and the peak over threshold, approaches for investigating extremes are asymptotically equivalent. 

Classical extreme value theory has been extended to deal with weakly correlated random variables \citep{leadbetteretal1983}, and adapted to analyse extremes of observables of chaotic dynamical systems. A detailed overview of this research field is provided by \citet{lucarini2016}. If one considers an Axiom A system, one obtains that extreme values of different classes of observables can be used to infer the properties of the stable and unstable manifold, including the possibility of estimating the Kaplan-Yorke dimension \citep{eckmann85}. Just as discussed above, adopting the chaotic hypothesis \citep{gallavotticohen1995}, such findings can be expected to apply for more general systems possessing high-dimensional chaos; see a detailed analysis in \citep{bodai_2017} and an accurate investigation in the case of a high-dimensional system (with O($10^3$) degrees of freedom) in \cite{galfietal2017}. Extreme Value Theory combined with the analysis of recurrences has proved very useful for providing a new framework for identifying the so-called weather patterns in actual climate data and in the outputs of climate models, and for interpreting their specific dynamical properties \citep{farandaetal2017,messorietal2017}. 

\subsection{Return Periods and Return Levels}
We compare the two methods of analysing rare events on a practical level, i.e. based on return periods and return levels. In case of LDT, we estimate the return periods $r$ of events exceeding the value $a$ using the general formula $r=\frac{1}{P(A_{n}>a)}=\frac{1}{1-P(A_{n}\le a)}$, where $P(A_{n}\le a)$ represents the cumulative distribution function of the large deviation law according to data. In case of the peak over threshold approach, the expected return levels can be written explicitly in terms of the Generalised Pareto parameters, which can be inferred using usual proven estimation methods, like maximum likelihood estimation \citep{coles2001} or L-moments \citep{hosking1990}. The level $y_r$ that is exceeded on average once every $r$-observations is called the $r$-observation return level and is the solution of $r=\frac{1}{P(Y>y)}$. One obtains $P(Y>y)$ from $H(y)-1=P(Y>y|Y>u)=\frac{P(Y>y)}{P(Y>u)}$, and consequently \citep{coles2001}:
\begin{equation}
  y_r=
  \begin{cases}
    u-\frac{\tilde{\sigma}}{\tilde\xi}\left[1-(\frac{1}{r P(Y>u)})^{-\tilde\xi}\right]       & \quad \mathrm{for } \ \tilde\xi \neq 0,\\
    u-\tilde{\sigma} \mathrm{ log } (\frac{1}{r P(Y>u)})                    & \quad \mathrm{for } \ \tilde\xi = 0.\\
  \end{cases}
  \label{eq:rlgpd}
\end{equation}

As an effect of serial correlations, the threshold exceedances can be organised in clusters. If an extreme value law does exist at all in this case, it is necessary to introduce the so-called extremal index - the inverse of the limiting mean cluster size - which has to be considered in the estimation of the Generalised Pareto parameters, with the exception of the shape parameter \citep{coles2001}. A widely adopted method to deal with correlated threshold excesses is to apply declustering, which basically aims to identify the maximum excess within each cluster and then to fit the Generalised Pareto distribution to the cluster maxima \citep{leadbetteretal1989,ferrosegers2003}. Here, since the operation of averaging reduces dramatically the effect of serial correlation, the peak over threshold approach can be applied in a straightforward way, similarly to the case of independent and identically distributed random variables.
  
\section{Model Description and Setup}
\label{sec:model}

We perform simulation with the Portable University Model of the Atmosphere (PUMA), which is a simplified spectral general circulation model developed at the University of Hamburg. PUMA has been used for the investigation of several atmospheric phenomena, like storm track dynamics or low frequency variability \citep{lunkeitetal1998,fraedrichetal2005a}, and has been even adapted to extra-terrestrial atmospheres \citep{griegeretal2004}. A recent study investigates the properties of the Lyapunov spectrum in PUMA, including large deviations of finite time Lyapunov Exponents \citep{decruzetal2018}. PUMA is the dry core of the Planet Simulator (PlaSim), which is a climate model of intermediate complexity \citep{fraedrichetal2005b,LFL2010}.

In the following, we summarise the model equations and the applied parametrisations. For a more detailed description of the model, please consult \citet{fraedrichetal2009}. As commonly done in atmospheric modelling, the physics of the model is fundamentally described by the primitive  equations for the atmosphere, which amount to a modification of the Navier-Stokes equation in a rotating frame of reference where the vertical acceleration of the fluid is constrained to be small compared to gravity \citep{klein2010}. These equations provide a good representation of the dynamics of the atmosphere for horizontal spatial scales larger than few tens of kms \citep{holton2004}.  Compared to a full atmospheric general circulation model, moist processes are omitted, and simple parametrisations are used to account for the effect of friction (Rayleigh friction), diabatic heating (Newtonian cooling), and diffusion. The Newtonian cooling and Rayleigh friction terms are such as that proposed by \citet{heldandsuarez1994} for the comparison of dynamical cores of general circulation models. The model equations allow for the conservation of momentum, mass, and energy. The prognostic equations for absolute vorticity $(\zeta+f)$, divergence $D$, temperature $T$, and surface pressure $p_s$ can be written by using spherical coordinates and the vertical $\sigma$-system as follows:

\begin{equation}
 \frac{\partial (\zeta+f)}{\partial t}=\frac{1}{1-\mu^2}\frac{\partial F_v}{\partial \lambda}-\frac{\partial F_u}{\partial \mu}-\frac{\zeta}{\tau_F}-K\nabla^8\zeta
 \label{eq:vor} 
\end{equation}

\begin{equation}
 \frac{\partial D}{\partial t}=\frac{1}{1-\mu^2}\frac{\partial F_u}{\partial \lambda}+\frac{\partial F_v}{\partial \mu}-\nabla^2\left(\frac{U^2+V^2}{2(1-\mu^2)}+\Phi+T_0\ln p_s\right)-\frac{D}{\tau_F}-K\nabla^8 D
 \label{eq:div} 
\end{equation}

\begin{equation}
 \frac{\partial T'}{\partial t}=-\frac{1}{1-\mu^2}\frac{\partial (UT')}{\partial \lambda}-\frac{\partial (VT')}{\partial \mu}+DT'-\dot{\sigma}\frac{\partial T}{\partial \sigma}+\kappa 
    \frac{T}{p}\omega+\frac{T_R-T}{\tau_R}-K\nabla^8T
 \label{eq:temp} 
\end{equation}

\begin{equation}
 \frac{\partial \ln p_s}{\partial t}=-\int_0^1 (D+\vec{V} \cdot \nabla \ln p_s) d\sigma
 \label{eq:spr} 
\end{equation}

with 
\begin{equation*}
 F_u=V(\zeta+f)-\dot{\sigma}\frac{\partial U}{\partial \sigma}-T'\frac{\partial \ln p_s}{\partial \lambda}
\end{equation*}

\begin{equation*}
 F_v=-U(\zeta+f)-\dot{\sigma}\frac{\partial V}{\partial \sigma}-T'(1-\mu^2)\frac{\partial \ln p_s}{\partial \mu}.
\end{equation*}
 
The variables and parameters used in Eq. (\ref{eq:vor}) -- (\ref{eq:spr}) are listed in Table \ref{tab:par}. 
\begin{table}[h]\small
  
  \caption{list of variables and parameters in puma, eq. (\ref{eq:vor}) -- (\ref{eq:spr}).}
  \centering
  \resizebox{.7\textwidth}{!}{\begin{tabular}{l l l}
    \toprule
    symbol									&value			& description \\
    \midrule
    $\zeta=\frac{\partial v}{\partial x}-\frac{\partial u}{\partial y}$		&			& relative vorticity \\
    $f$										&			& Coriolis parameter \\
    $D=\frac{\partial u}{\partial x}+\frac{\partial v}{\partial y}$		&			& horizontal divergence \\
    $T$										&			& temperature \\
    $T_0$									&$250 \unit{K}$		& reference temperature \\
    $T'=T-T_0$									&			& temperature deviation from $T_0$ \\
    $p$										&			& pressure \\
    $p_s$									&			& surface pressure \\
    $\sigma=p/p_s$								&			& vertical coordinate \\
    $U=u \cos{\phi}$								&			& zonal velocity in spherical coordinates \\
    $V=v \cos{\phi}$								&			& meridional velocity in spherical coordinates \\
    $\vec{V}$									&			& horizontal velocity with components $U$ and $U$ \\
    $t$										&			& time \\
    $\phi$									&			& latitude \\
    $\mu$									&			& $\sin{\phi}$ \\
    $\lambda$									&			& longitude \\
    $\phi$									&			& geopotential \\
    $\omega=dp/dt$								&			& vertical velocity in $p$-system \\
    $\dot{\sigma}=d\sigma/dt$							&			& vertical velocity in $\sigma$-system \\
    $\tau_F$									&			& time scale for Rayleigh friction \\
    $K$										&			& hyperdiffusion coefficient \\
    $\tau_R$									&			& time scale for Newtonian cooling \\
    $T_R$									&			& restoration temperature \\
    $\kappa$									&$0.286$		& adiabatic coefficient \\
    \bottomrule
  \end{tabular}}
  \label{tab:par}
\end{table}

The horizontal representation of the prognostic model variables is given by a series of spherical harmonics, which are integrated in time by a semi-implicit time differencing scheme \citep{hoskinsandsimons1975}. The linear contributions in the prognostic equations are computed in spectral space, the non-linear contributions in grid point space. The horizontal resolution is defined by triangular truncation. The vertical discretization is based on finite differences on equally spaced $\sigma$-levels. The vertical velocity is set to $0$ at the upper ($\sigma=0$) and lower ($\sigma=1$) boundaries.

A Rayleigh damping of horizontal velocities with time scale $\tau_F$ accounts for the effect of boundary layer friction in the lowest levels. $\tau_F=0.6$ days at $\sigma=0.95$ (the vertical level nearest to surface), and $\tau_F=1.65$ days at $\sigma=0.85$. For higher levels no friction is considered, i.e $\tau_F=\infty$.
The effect of non resolved processes on the energy and enstrophy cascade is represented by hyperdiffusion ($\sim \nabla^{2 h}$). The hyperdiffusion coefficient $K$ is such that provides a maximal damping of the shortest waves, and has no effect on the mean state (wave number $0$). The integer exponent $h=4$ leads to an additional damping of short waves. The diffusion time scale for the shortest wave is $1/4$ days. 
The diabatic heating (cooling) is parameterized by a Newtonian cooling term. This forces the relaxation of the model temperature to a so-called radiative-convective \textit{equilibrium} state specified by the restoration temperature $T_R$, which depends only on vertical level and latitude.
\begin{equation}
 T_R(\phi,\sigma)=T_R(\sigma)+f(\sigma)T_R(\phi)
\end{equation}
$T_R(\phi)$ describes the meridional form of the restoration temperature, whereas $f(\sigma)$ accounts for the vertical changes in this meridional profile:
\begin{equation}
 T_R(\phi)=(\Delta T_R)_{NS}\frac{\sin \phi}{2}-(\Delta T_R)_{EP}(\sin^2{\phi}-\frac{1}{3}),
\end{equation}
where $(\Delta T_R)_{NS}$ is the temperature difference between the North and South poles, and $(\Delta T_R)_{EP}$ represents the equator-to-pole temperature difference. 
The meridional temperature gradient decreases with height in the troposphere, $f(\sigma)=\sin (0.5\ \pi (\sigma-\sigma_{tp})/(1-\sigma_{tp}))$ for $\sigma \ge \sigma_{tp}$, and vanishes at the tropopause, $f(\sigma)=0$ for $\sigma < \sigma_{tp}$, where $\sigma_{tp}$ is the height of the tropopause. $T_R(\sigma)$ describes the vertical profile of the restoration temperature:
\begin{equation}
 T_R(\sigma)=(T_R)_{s}-L z_{tp}+\sqrt{{\left[\frac{L}{2}\big ( z_{tp}-z(\sigma)\big)\right]}^2+S^2}+\frac{L}{2}\big(z_{tp}-z(\sigma)\big),
\end{equation}
with: restoration temperature at the surface, $(T_R)_s=288 \unit{K}$; moist adiabatic lapse rate, $L=6.5 \unit{K/km}$; global constant height of the tropopause, $z_{tp}=12 \unit{km}$; geometric height $z$. $S$ allows for a smoothing of the temperature profile at the tropopause. In case of 10 vertical levels $l$, the time scale of the Newtonian cooling $\tau_R$ is $2.5$ days in the lowest level at $l=10$, and 7.5 days at $l=9$. $\tau_R$ continues to increase monotonically with height until the upper 3 levels, where it is set to 30 days.

We run the model in a simple symmetric setting (usually referred to as aqua-planet), i.e without orography. We remove the annual and diurnal cycle, and use a symmetric forcing with respect to the Equator, $(\Delta T_R)_{NS}=0$. We set the equator-to-pole temperature difference $(\Delta T)_{EP}$ to $90 \unit{K}$, thus creating a baroclinically more unstable atmospheric state than in the standard setting with $(\Delta T)_{EP}=70 \unit{K}$. We run the model with constant forcing in time using a time step of 30 minutes. The horizontal resolution is T$42$ (triangular spectral truncation with 42 zonal waves), and the vertical resolution consists of 10 levels. The length of the simulations is $10^4$ years, excluding a transient of $5$ years, which are discarded to reach steady state. We consider for our analysis the air temperature in the lowest vertical level at $960 \unit{hPa}$, with daily output. The spectral temperature variable is transformed during the post-processing into grid point space consisting of a $65 \times 128$ equidistant latitude-longitude grid.

Using the same model settings as above, but with a lower Equator-to-Pole temperature difference, \citet{decruzetal2018} estimate a Kaplan-York dimension $D_{KY}$ of $187$ and a number of positive Lyapunov exponents of $68$ for $(\Delta T)_{EP}=60 \unit{K}$. In this study, $(\Delta T)_{EP}=90 \unit{K}$, thus the model atmosphere is baroclinically substantially more unstable than in the mentioned study. Thus, to provide a rough estimation, $D_{KY}>200$ and the number of positive Lyapunov exponents $>80$ in our system. Consequently, we expect for this set-up a very high dimensional chaos, which fulfils the chaotic hypothesis, as shown also by the fast decay of auto-correlations in Fig.~\ref{fig:gen}c,d below. As a result, we expect that the outputs of our model can be analysed using extreme value theory and LDT, as discussed above. Nonetheless, it is \textit{a priori} unclear whether the asymptotic result can be clearly detected at finite size given the length of our numerical integrations. Note that in \citet{decruzetal2018} it was shown that the finite time Lyapunov exponents obey a large deviation law.

\section{Results}
\label{sec:results}

Before discussing our main results related to the large deviations of temperature, it is useful to have a general picture about the properties of the simulated temperature field at $960 \unit{hPa}$ (i.e. close to the surface). For the analysis of temporal, zonal, and spatio-temporal large deviations, we select three latitudes: $60\degree$, $46\degree$, and $30\degree$. 
We focus on the mid-latitudes because it is the region of the atmosphere with the strongest turbulence, so that we expect that the corresponding observables should behave in agreement with the chaotic hypothesis; see discussion in \cite{galfietal2017}. We remark that the inclusion of moist processes, of more comprehensive paramerisations, and less idealised boundary conditions would greatly increase chaotic processes, and higher horizontal and vertical resolutions would lead to substantially stronger turbulence in the tropical belt. 

In the considered setting, the two hemispheres have identical statistical properties; additionally, the two hemispheres are weakly coupled, broadly as a result of chosen boundary conditions, of the lack of seasonal cycle, and of the conservation law for potential vorticity. Therefore, we can treat the time series coming from the two hemispheres as separate realisations of the same dynamical process. In the following, we provide first a qualitative comparison of temporal and spatial features of the temperature field, and then we quantify the persistence in time and space based on the integrated auto-correlation. We point out that we perform the analysis in a Eulerian framework, corresponding to our objective of studying persistent temperature extremes from a spatially fixed point of view: this provides the most relevant information for the specific problem - investigation of persistent temperature anomalies - we have in mind.

\begin{figure}[ht!]
  \centering
  \includegraphics[width=1\textwidth]{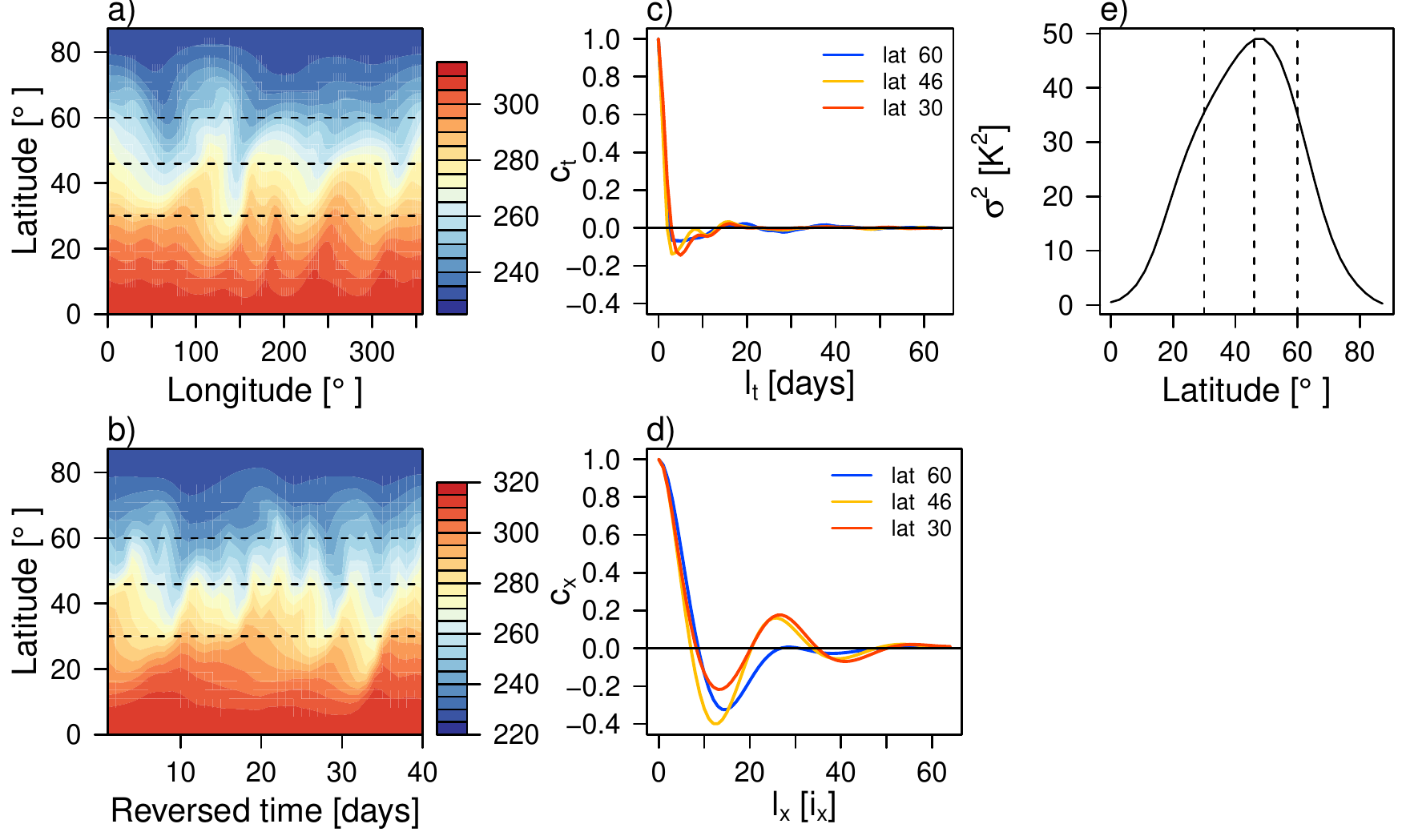}
  \caption{General properties of the temperature field at 960 hPa. a) Temperature values $T(x,y,t^*)$ as function of longitude $x$ and latitude $y$ at one selected time point $t^*$. b) Temperature values $T(x^*,y,t)$ as function of latitude $y$ and of time $t$ at one selected longitude $x^*$ (the $x$-axis is backward in time). c) Temporal and d) zonal auto-correlation functions according to (\ref{eq:timcor}) and (\ref{eq:zoncor}) for the selected latitudes (different colours according to the legend). e) Variance $\sigma^2$ of near-surface temperature according to (\ref{eq:var}). The dashed lines in a), b), and e) mark the selected latitudes. }
  \label{fig:gen}
\end{figure}

Figure~\ref{fig:gen}a illustrates the temperature field $T(x,y,t^*)$ as function of longitude $x$ and latitude $y$ at one selected time point $t^*$, whereas Fig.~\ref{fig:gen}b represents the temperature field $T(x^*,y,t)$ as function of latitude $y$ and of time $t$ at one selected longitude $x^*$. Qualitatively similar figures would be obtained for different values of $t^*$ and of $x^*$, respectively. 
Note that, to facilitate the comparison between space and time, the $x$-axis in Fig.~\ref{fig:gen}b is backward in time according to the prevailing eastward zonal wind at mid-latitudes \citep{holton2004}. Additionally, the range of the $x$-axis in Fig.~\ref{fig:gen}b is the same as in Fig.~\ref{fig:gen}a once we rescale the time axis according to the scale velocity $U_\tau$ introduced below (computed for $46\degree$), which weights the decay of correlation in space at fixed time and in time at fixed location. Comparing these two figures we realise that by cutting across time or across longitudes we obtain very similar wavy patterns, which is non-surprising since the forcing is invariant in time and along a latitudinal band. 

While this result would be trivial when observing a periodic or quasi-periodic signal, we need to consider here that the dynamics of the atmosphere features a non-trivial mixture of wave, turbulence, and particles \citep{ghilandrobertson2002}, so that we need to look at this space-time similarity from a statistical point of view. According to this, we have that, at a given latitude $y^*$, the temporal series $T(x^*,y^*,t)$ and the zonal series $T(x,y^*,t^*)$ are sampled from two similarly distributed random processes, given  the condition of steady state and the discrete symmetry with respect to translation along latitudes.  

The main difference between $T(x^*,y^*,t)$ and $T(x,y^*,t^*)$ is related to distinct temporal and spatial characteristic scales, i.e. to temporal or spatial correlations. At mid-latitudes, cyclones have a typical temporal scale of $\approx 1$ day and a characteristic spatial scale of about $1000 \unit{km}$ \citep{holton2004}. Obviously, these scales are relevant when we try to obtain a large deviation principle, thus it is very important to find an adequate metric to describe them. We quantify the typical temporal and zonal scales based on the integrated auto-correlation, as explained in Sec.~\ref{sec:theory}.

We calculate the auto-correlations of the temporal and zonal series at a selected latitude $y^*$, based on which we obtain later the integrated temporal and zonal auto-correlations. For this, we use 1000 years of our simulation out of a total of 10000 years, as this proves to be more than enough to reach robust estimates. As described in Sec.~\ref{sec:theory}, the auto-correlation is defined as the ratio between the auto-covariance $C(l)$ at lag $l$ and the variance $\sigma^2$: $c(l)=C(l)/C(0)=C(l)/\sigma^2$. To obtain better auto-correlation estimates, we calculate the spatio-temporal mean and variance at each $y^*$, and use these estimates for the computation of both temporal and zonal auto-correlations: 
\begin{equation}
\mu=\frac{1}{N_tN_x}\sum_{j=1}^{N_t}\sum_{i=1}^{N_x}T(i,y^*,j)
\end{equation}
and 
\begin{equation}
\sigma^2=\frac{1}{N_tN_x-1}\sum_{j=1}^{N_t}\sum_{i=1}^{N_x}\left(T(i,y^*,j)-\mu\right)^2,
\label{eq:var}
\end{equation}
where $N_t=3.6\times 10^5$ is the number of considered points in time (daily data), and $N_x=128$ is the number of grid points in the zonal direction. This is reasonable considering the symmetries in our system in time and along latitudinal circles. The subscripts $t$ and $x$ refer to time and to the zonal dimension, also in what follows.

In the case of the temporal series $T(x^*,y^*,t)$,  we calculate the auto-covariance at one selected longitude $x^*$. This estimate is independent of $x^*$, thus it is unimportant which longitude we choose. We have:

\begin{equation}\label{eq:timcor}
 c_t(l_t)=\frac{1}{\sigma^2}\frac{1}{N_t}\sum_{i=1}^{N_t-l_t}\left(T(x^*,y^*,i)-\mu\right)\left(T(x^*,y^*,i+l_t)-\mu\right)
\end{equation}

The length of the zonal series $T(x,y^*,t^*)$, however, is too short to obtain reliable auto-correlation estimates. The number of grid points along the zonal dimension is only $128$. Together with such a restriction related to the \textit{size} of the Earth, there is another one related to the \textit{shape} of the Earth. In fact,  we have to reduce the maximum lag to $N_x/2=64$ because at larger lags the correlations start to increase again due to the periodicity along a latitudinal circle. To increase the robustness of our estimate, we first calculate the lagged zonal auto-correlation coefficients at each time point and then we take the average over time:

\begin{equation}\label{eq:zoncor}
 c_x(l_x)=\frac{1}{\sigma^2}\frac{1}{N_x N_t}\sum_{j=1}^{N_t}\sum_{i=1}^{N_x-l_x}\left(T(y^*,i,j)-\mu\right)\left(T(y^*,i+l_x,j)-\mu\right).
\end{equation}

Figure~\ref{fig:gen}c shows the temporal auto-correlation coefficient as function of the temporal lag in units of days, whereas Fig.~\ref{fig:gen}d illustrates the zonal auto-correlation coefficient as function of the spatial lag expressed as longitude indexes $i_x=0,1,2,...$. Both temporal and spatial auto-correlations decay to zero, meaning that two temperature values, which are far away from each other in time or in space are independent for all practical purposes. We finally estimate the integrated temporal and zonal auto-correlations by taking the sum of the auto-correlation coefficients until the maximum lag $l_t=l_x=64$. Note that we use the same temporal and zonal maximum lags for consistency reasons. The temporal integrated auto-correlation can be obtained also for larger maximum lags, but this changes only negligibly the estimate value because the decay to 0 is relatively fast. We define:
\begin{subequations}
\label{eq:tauint}
\begin{align}
 \tau_t=1+2\sum_{l_t=1}^{64} c_t(l_t),\\
 \tau_x=1+2\sum_{l_x=1}^{64} c_x(l_x).
\end{align}
\end{subequations}
$\tau_t$ is $1.32$ at $60\degree$, $1.05$ at $46\degree$, and $1.61$ at $30\degree$ (in units of time steps, which are equivalent to days), whereas $\tau_x$ is $3.26$ at $60\degree$, $3.54$ at $46\degree$, and $7.68$ at $30\degree$ (in units of gridpoints, which correspond to 391 km at $60\degree$, 732 km at $46\degree$, and 1292 km at $30\degree$). We define $\tau_t$ and $\tau_x$ in a non-dimensional form (i.e. as number of time units or zonal data points) to facilitate the comparison of temporal and spatial persistence based on the resolution of our data, and because in this form we can use them directly for scaling the rate function, as we show below. 

We define the scale velocity $U_\tau:=\frac{\tau_x \delta_x}{\tau_t \delta_t}$, where $\delta_t$ is the time step of $1 \unit{day}$ and $\delta_x$ is the latitude dependent grid spacing. From a statistical point of view, $U_\tau$ is the ratio between spatial and temporal persistence.\footnote{A straightforward connection between spatial and temporal typical scales would be the phase speed of patterns, which is represented on synoptic scales at mid-latitudes by the phase speed of Rossby waves. As discussed in the introduction, this phase speed depends on the wave length, thus, unlike $U_\tau$, it is not a general property of the flow at a certain latitude.} From a geometrical/dynamical point of view, $U_\tau$ represents the ratio between spatial and temporal typical scales. Thus, $U_\tau$ is a measure for the anisotropy between space and time. At $60\degree$ $U_\tau=4.25 \unit{ms}^{-1}$  and at $46\degree$ $U_\tau=8.47 \unit{ms}^{-1}$. For these latitudes, the scale velocity $U_\tau$ is in good agreement with the mean zonal velocity at 960 hPa $\overline{[U]}$, which is $3.6 \unit{ms}^{-1}$ at $60\degree$, and $6 \unit{ms}^{-1}$ at $46\degree$. This is hardly surprising as, to a first approximation, the turbulent structures are advected by the mean flow.

The agreement is lost when looking at $30\degree$, the boundary of the mid-latitude baroclinic zone, for which the qualitative description given above applies. As we approach the equator, the atmospheric dynamics has a much lower degree of chaoticity with respect to the mid-latitudes, unless we look at convective scales, which are not resolved at all in this model. The spatial persistence is strongly enhanced (see also Fig.~\ref{fig:gen}a,b), as a result of the dominance of larger structures associated to the downdraft of the Hadley cell rather than synoptic disturbances associated to mid-latitude weather systems advected by the prevailing westerlies. In this case we find $U_\tau=14.95 \unit{ms}^{-1}$, while $\overline{[U]}$ is $-3.4 \unit{ms}^{-1}$, which indicates prevailing easterly flow, a clear signature of tropical dynamics. 

Figure~\ref{fig:gen}e emphasises that the near-surface temperature experiences the largest variance near latitude $46 \degree$, as a result of the very strong baroclinic instability associated to mid-latitude weather patterns. This latitude also corresponds to the mean-position of the jet, the localised region where the speed of the upper-level winds are maximum \citep{holton2004}.

\begin{figure}[ht!]
  \centering
  \includegraphics[width=.7\textwidth]{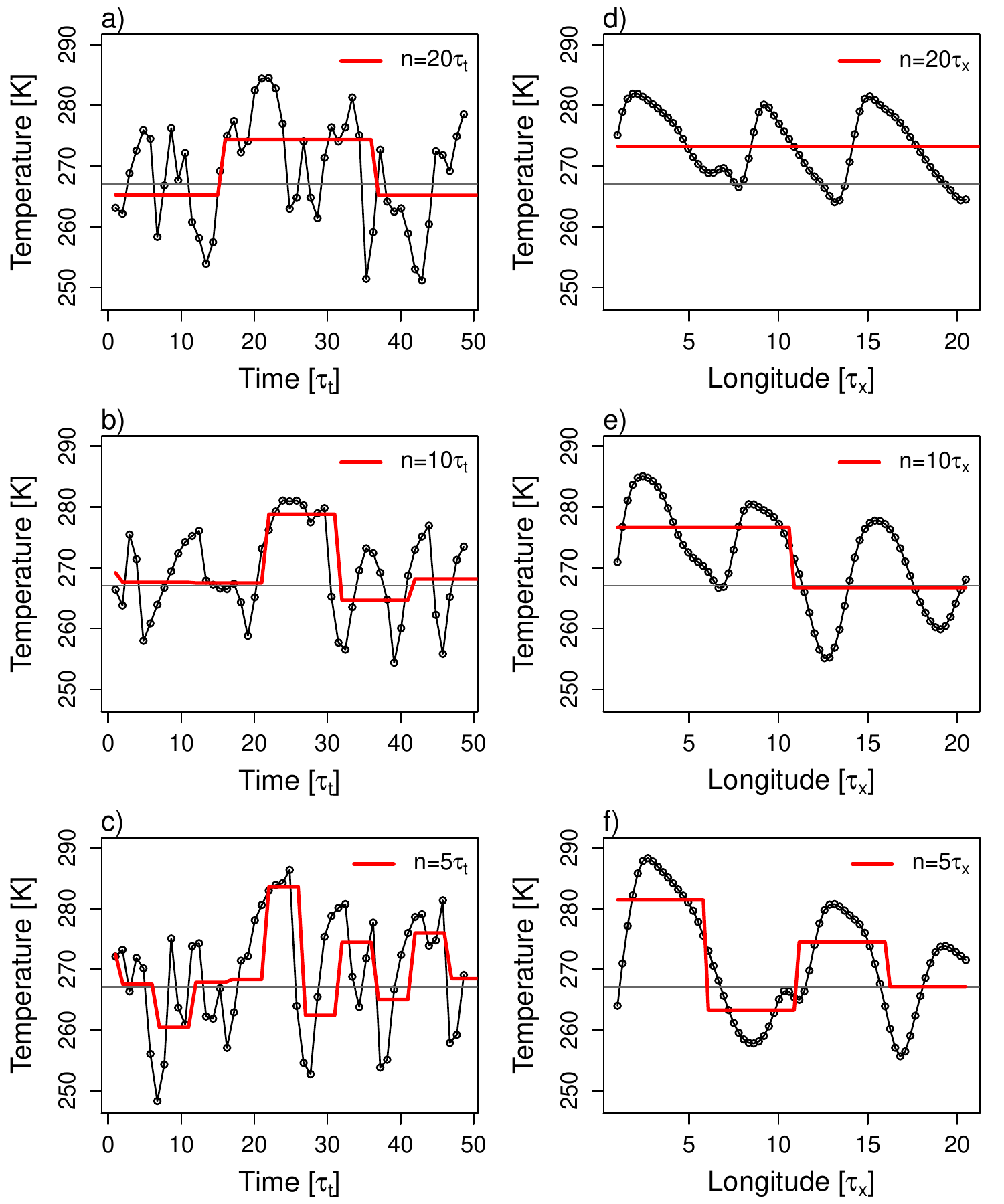}
  \caption{Relationship between persistent events and large fluctuations of the coarse-grained fields. a) Time series (black line) of near-surface temperature at $46\degree$ in the case of a large event of the coarse grained time series (red line) with averaging window of $20\tau_t$; $x$-axis in units of $\tau_t=1.05$ days.  b) Same as a), for averaging window of $10\tau_t$. c) Same as a), for averaging window of $5\tau_t$.  d) Zonal series (black line) of surface temperature at $46\degree$ in the case of a large event of the coarse grained zonal series (red line) with averaging window of $20\tau_x$; $x$-axis in units of $\tau_x=3.54$ grid points. e) Same as f), for averaging window of $10\tau_x$. f) Same as d), for averaging window of $5\tau_x$. In all panels the grey horizontal line represents the long term and longitudinal average.} 
  \label{fig:gen2}
\end{figure}

Before continuing with the description of the temporal and spatial large deviations, we briefly discuss the connection between high values of coarse grained temperatures and long individual events where the temperature readings are persistently above the long-term average, discussed already in Sec.~\ref{sec:intro}. Fig.~\ref{fig:gen2}a,b,c show three short temporal series at latitude $46\degree$  together with the corresponding series of the coarse grained quantities where averages are computed using block lengths of $20 \tau_t$, $10 \tau_t$, and $5 \tau_t$, respectively. The three short time series have been specifically chosen because they feature a large fluctuation in the coarse grained quantity. Fig.~\ref{fig:gen2}d,e,f show the same in the case of the zonal fields. The main finding is that up to moderately long averaging windows of about  $10\tau_t$ (or $5\tau_x$ for spatial averages) it is possible to link large fluctuations with individual persistent events. When a coarser graining is considered, using a window of $20\tau_t$ for time averages and  $10\tau_x$ or $20\tau_x$  for spatial ones, thus going in the direction of the regime of the large deviations discussed below, we do not have such a one-to-one identification. Instead, large ultralong fluctuations are related to the occurrence of subsequent moderately long persistent features.

\subsection{The Link between Temporal and Zonal Large Deviations}
\label{subsec:link}

At this point, we turn our attention to the estimation of the temporal and zonal rate functions. For this, we first have to obtain sequences of temporal and zonal averages for increasing lengths of averaging blocks $n_t$ and $n_x$, for which we use the total length of our simulation of $10000$ years.

\begin{subequations}\label{eq:nseq}
\begin{align}
 A_{n_t}=\frac{1}{n_t}\sum_{i=1}^{n_t} T(x^*,y^*,t=i)\label{eq:ntseq},\\
 A_{n_x}=\frac{1}{n_x}\sum_{i=1}^{n_x} T(x=i,y^*,t^*)\label{eq:nxseq}.
\end{align}
\end{subequations}

The lengths of temporal averaging blocks are chosen to be multiples of $\tau_t$: $n_t=5\tau_t,10\tau_t,...,40\tau_t$. Similarly, the lengths of zonal averaging blocks are multiples of $\tau_x$, but in this case the largest possible multiple $m$ is limited due to the size and shape of the Earth, as mentioned above: $n_x=5\tau_x,10\tau_x,...,m\tau_x$. $m=20$ in case of latitudes $60\degree$ and $46\degree$, whereas $m=10$ in case of latitude $30\degree$. To increase the number of averaged values for the computation of the temporal rate functions, we lump together the temporal averages from every 25th longitude along a latitudinal circle. Since $\tau_x\ll 25$, these temporal sequences can be treated as independent realisations.\footnote{We remark that the aim here is to obtain better \textit{temporal} large deviation estimates. For this purpose it is important to exclude \textit{spatial correlation}s, thus the time series at the chosen latitudes should not be correlated.} In case of zonal averaging, we take one averaged value in space from every 10th point along the time axis, which we consider to be independent realisations as well.\footnote{For \textit{spatial} averaging, we use a similar argument as for temporal averages. The aim is to exclude, at this point, the effect of \textit{temporal correlations}.} Such an assumption is reasonable because the integrated temporal auto-correlation of zonal averages is much lower then 10, even for the largest $n_x$ (as shown later in Fig.~\ref{fig:stac}). We obtain for each value of $n_t$ and $n_x$ estimates of the rate functions, after using the re-normalising factors given by $1/\tau_t$ or $1/\tau_x$, respectively: 

\begin{subequations}\label{eq:rfest}
\begin{align}
 \tilde{I}_{n_t}(a)=-\frac{\ln p(A_{n_t}=a)}{n_t}\tau_t \label{eq:rfestt},\\
 \tilde{I}_{n_x}(a)=-\frac{\ln p(A_{n_x}=a)}{n_x}\tau_x \label{eq:rfestx},
\end{align}
\end{subequations}

where $p(A_{n_t}=a)$ and $p(A_{n_x}=a)$ represent empirical estimates of the pdf's of the temporally and zonally averaged sequences. Due to the re-normalisation, the logarithm of the probabilities is scaled by $n_t/\tau_t$ or $n_x/\tau_x$, i.e. by the  number of uncorrelated, instead by the total number of, data in an averaging block. Thus, we eliminate the effect of correlations. 

\begin{figure}[ht!]
  \centering
  \includegraphics[width=1\textwidth]{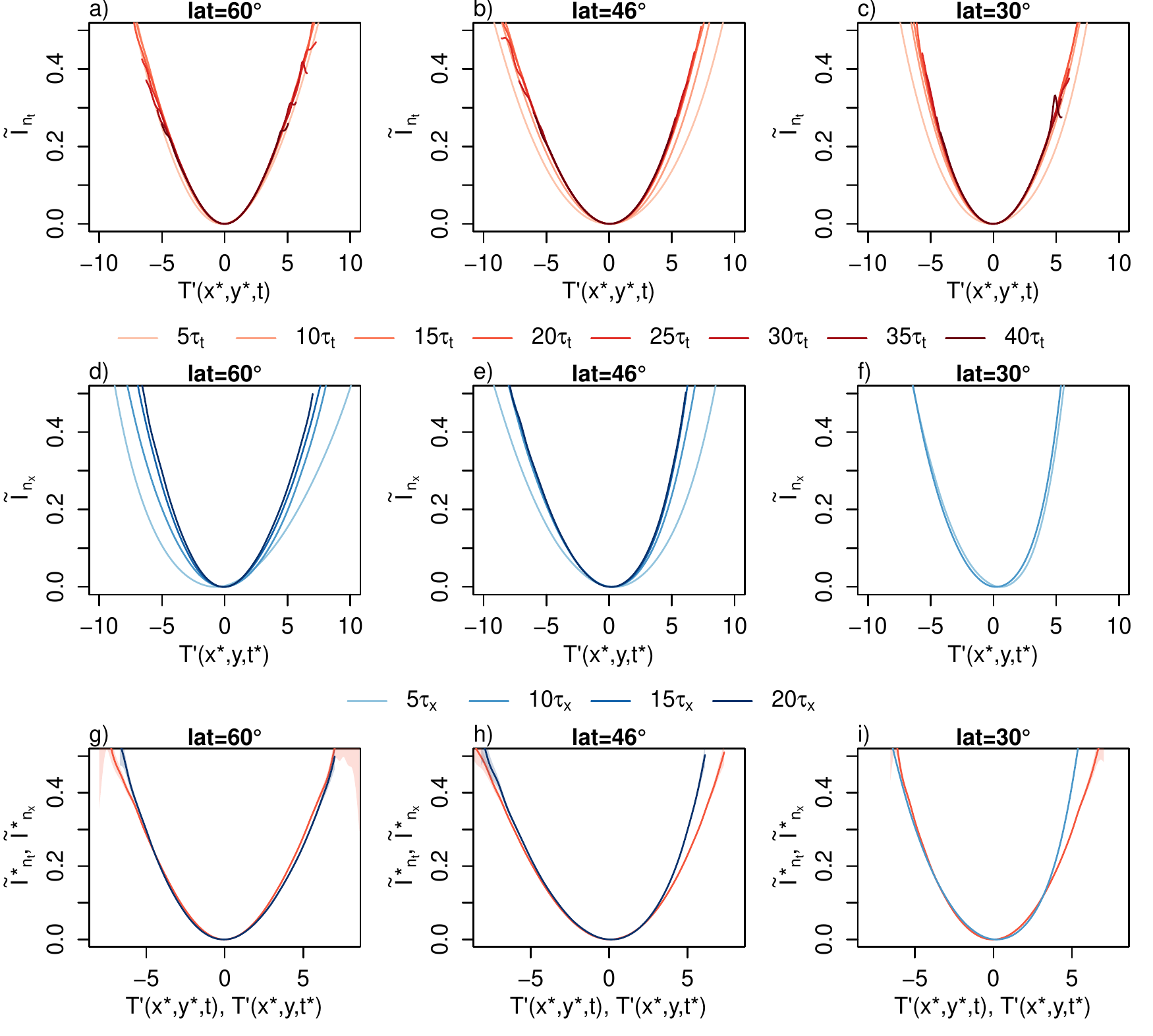}
  \caption{a) -- c) Temporal re-normalised rate function estimates $\tilde{I}_{n_t}$ and d) -- f) zonal re-normalised rate function estimates $\tilde{I}_{n_x}$ for the three considered latitudes and for increasing averaging lengths $n_t$ and $n_x$ according to the different colours (see legend). g) -- i) Best estimates of the temporal (red) and zonal (blue) re-normalised rate functions. All estimates are shifted vertically so that their minimum is at 0. $T'=T-\mu$ represents temperature fluctuations around the mean.} 
  \label{fig:rf1}
\end{figure}

Figure \ref{fig:rf1} shows $\tilde{I}_{n_t}$ (a--c) and $\tilde{I}_{n_x}$ (d--f) for every $n_t$ and $n_x$. As a side note, we remark that in every figure below the shown re-normalised rate function estimates are shifted vertically so that their minimum is at 0. In case of the temporal rate functions, it is clear that for $n_t\ge 20\tau_t$ the estimates $\tilde{I}_{n_t}$ do not change in shape by further increase in $n_t$, meaning that we obtain stable and reliable estimates, i.e. there is a proof in our data for a large deviation principle in time. We also notice that the range of $A_{n_t}$ values becomes narrower as $n_t$ increases as an effect of averaging, which reduces the amount of available data. Thus, we obtain our best estimate at an optimal averaging block length $n_t^*$ which is large enough to allow for the convergence of rate function estimates, but is in the same time small enough so that the range of $A_{n_t}$ is not too narrow, i.e. $n_t^*=\min(n_t;\tilde{I}_{n_t}\approx I_{n_t})$. We choose the same optimal averaging length for all three latitudes: $n_t^*=20\tau_t$; although in case of latitudes $60\degree$ and $30\degree$, $\tilde{I}_{n_t=10\tau_t}$ seems to be already a good estimate for the asymptotic $I_{n_t}$. Comparing the re-normalised rate function estimates at the selected latitudes, we realise that the rate function has smaller curvature at latitude $46 \degree$ than at $60 \degree$ and $30 \degree$. This is not a trivial consequence of the larger variability of the system in the middle of the considered domain, as mentioned above and shown in Fig.~\ref{fig:gen}e, because we are considering here averages of fluctuations. 

In case of the zonal rate functions, we first notice that the largest $n_x$ seems to be too small for a clear-cut convergence. In other words, the length of a latitudinal circle is not long enough to clearly obtain a large deviation limit. However, the dependence of $\tilde{I}_{n_x}$ on $n_x$ seems to decrease as $n_x$ is increasing, thus we choose the largest possible $n_x$ as the optimal zonal averaging length $n_x^*=\max(n_x)$. $n_x^*=20\tau_x$ in case of latitudes $60\degree$ and $46\degree$, whereas in case of latitude $30\degree$, $n_x^*$ is only $10\tau_x$ because of stronger zonal auto-correlations. 

The best estimates of the temporal and zonal re-normalised rate functions $\tilde{I}^*_{n_t}=\tilde{I}_{n_t=n_t^*}$ and $\tilde{I}_{n_x}^*= \tilde{I}_{n_x=n_x^*}$ are shown again in Fig.~\ref{fig:rf1}g,h,i. The shading represents the 95\% confidence intervals of 2000 nonparametric ordinary bootstrap estimates based on the normal distribution (functions \texttt{boot} and \texttt{boot.ci} of the \textsf{R} package \texttt{boot}, \citealp{boot1997,boot2017}). We notice that $\tilde{I}^*_{n_t}~\approx~\tilde{I}_{n_x}^*$. The equivalence is very good in case the of latitude $60 \degree$ and in the case of negative anomalies at latitudes $46 \degree$ and $30 \degree$. We also notice some differences for positive anomalies at latitudes $46 \degree$ and $30 \degree$, with larger differences at $30 \degree$. At this later latitude, however, it has to be considered that the maximum possible zonal averaging length is $10\tau_x$, whereas in the other cases it is $20\tau_x$. We assume that the differences between the temporal and zonal re-normalised rate function estimates have to do with the fact that $n_x^*$ is not large enough to estimate the rate function properly. Larger values of $n_x$ are needed to overcome the enhanced skewness in the distribution of zonal averages as effect of spatial correlations, however this is impossible due to limitations coming from the size and shape of the Earth. These findings have correspondence with the large value of  $U_\tau$ at this latitude, defining the anisotropy between space and time. While the temporal rate function can be estimated reliably at a relatively small $n_t$, the estimation of the zonal one is a much more difficult task.

However, the main message of Fig.~\ref{fig:rf1} is that the temporal and zonal re-normalised rate functions seem to be equal, $I_{n_t}=I_{n_x}$, if the probability of averages is scaled by the number of uncorrelated data in an averaging block, $n_t/\tau_t$ or $n_x/\tau_x$, as explained above. In other words, there is a link connecting temporal and spatial large deviations or averages, due to the existence of a \textit{universal function} $I_n$; universal in the sense that it represents large deviations in both dimensions: time and space.

Obviously, based on the large deviation principle in time or in space, one cannot characterise persistent temporal or spatial events, because the limit law starts to act on larges scales, where persistence is lost and universality emerges. However, one can capture persistent space-time events by averaging in both dimension, space and time. To achieve this, it is important that the spatial averaging length is not too small but not too large either, as we show in the following.

\subsection{Spatio-temporal Large Deviations}
\label{subsec:zontemp}

We consider temporal sequences of zonally averaged observables over averaging lengths $n_x=1\tau_x,5\tau_x,10\tau_x,20\tau_x$, and then average each sequence in time for increasing averaging lengths $\hat{n}_{t}=1\hat{\tau}_{t},5\hat{\tau}_{t},10\hat{\tau}_{t},15\hat{\tau}_{t},...40\hat{\tau}_{t}$. The notation $\hat{ }$ is meant to indicate that we average in space and \textit{additionally} in time, and $\hat{\tau}_{t}$ is the decorrelation time of the spatially averaged observable. By considering several $n_x$ values, we choose the spatial scale at which we analyse the large deviations in time.  The spatio-temporal averages are computed as:

\begin{equation}
 A_{n_x,\hat{n}_t}=\frac{1}{\hat{n}_{t}}\sum_{j=1}^{\hat{n}_{t}}\frac{1}{n_x}\sum_{i=1}^{n_x}T(i,y^*,j)=\frac{1}{\hat{n}_{t}}\sum_{j=1}^{\hat{n}_{t}} A_{n_x}(j).
\end{equation} 

\begin{figure}[ht!]
  \centering
  \includegraphics[width=.5\textwidth]{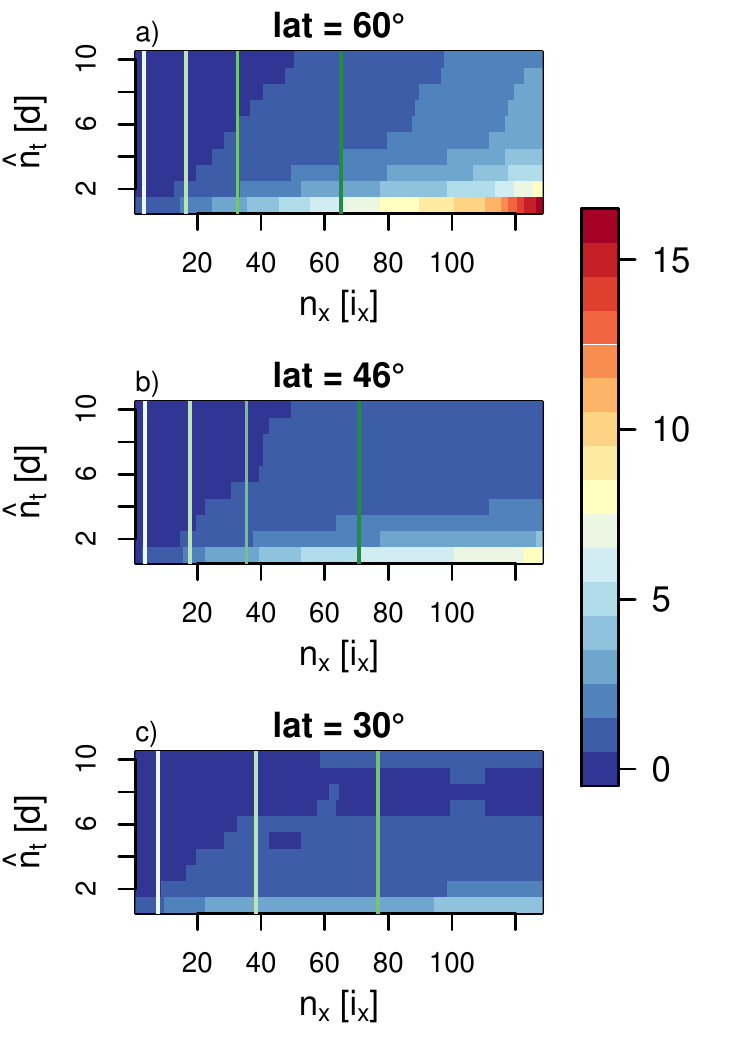}
  \caption{Integrated temporal auto-correlation of spatio-temporal averages $\hat{\tau}_t$ for the selected latitudes as function of zonal and temporal averaging lengths. The vertical lines mark zonal averaging lengths, corresponding to the multiples of $\tau_x$: $1\tau_x,5\tau_x,10\tau_x,20\tau_x$ (from white to green).} 
  \label{fig:stac}
\end{figure}

Similarly to the previous cases, also in case of spatio-temporal averages, we have to take into account the strength of auto-correlations if we pursue to compare the spatio-temporal rate functions with the temporal and zonal ones. We estimate the integrated temporal auto-correlation $\hat{\tau}_t$ of spatio-temporal averages analogously to $\tau_t$ or $\tau_x$, but, in order to assure the stability of $\hat{\tau}_t$, we choose a higher maximum lag of 120 days because the auto-correlation in time of zonal averages has a slower decay compared to the one of unaveraged temporal or zonal observables. Fig.~\ref{fig:stac} shows $\hat{\tau}_t$ as function of zonal averaging length $n_x$ and temporal averaging length $\hat{n}_t$. The temporal auto-correlations of the spatio-temporal observables are increasing with $n_x$ and decreasing with $\hat{n}_t$. The increase with $n_x$, on the one hand, can be explained by the connection between temporal and spatial scales. Large events in space are long lasting events in time, as discussed in Sec.~\ref{sec:intro}.\footnote{This statement is justified by the properties of Rossby wave propagation in the mid-latitudes as mentioned in the introduction. Accordingly, large enough waves can become stationary and lead to persistent temperature extremes.}
The decrease of the temporal integrated auto-correlation with $\hat{n}_t$, on the other hand, can be explained by the increase of the number of uncorrelated events with respect to the number of correlated events in an averaging block as a consequence of increasing the block length. This is automatically the case for large averaging blocks when correlations are finite, and is crucial for the applicability of the block averaging method. The different behaviour with $n_x$ and $\hat{n}_t$, however, has to do mainly with the discrepancies in the temporal resolution of the newly obtained averages. While, in the case of zonal averaging, the temporal resolution remains one day, in the case of additional averaging in time, the temporal resolution decreases with $\hat{n}_t$, thus the temporal auto-correlation lag increases. However, this is not a problem for our analysis since we are interested in the correlations of the averaged observables measured in number of averaged data. A stronger increase of $\hat{\tau}_t$ at the ``end'' of the channel underlines the above discussed effect of averaging along a latitudinal circle. At the zonal ``end'' of the channel, the temperature values are strongly correlated with the ones at the ``beginning'' of the channel.

The dependence of $\hat{\tau}_t$ on the zonal and temporal averaging lengths is qualitatively similar for the chosen latitudes if one considers $n_x$ in units of $\tau_x$ (along the vertical lines in Fig.~\ref{fig:stac} with same colours). As we proceed from South to North, the auto-correlations of the zonally averaged observables are becoming stronger. This is, however, mostly due to the decreasing distance between the longitudes leading to stronger correlated temperature values at neighbouring longitudes.

Estimates of spatio-temporal re-normalised rate functions are then computed for each $n_x$ and $\hat{n}_{t}$ as:

\begin{equation}
 \tilde{I}_{n_x,\hat{n}_{t}}(a)=-\frac{\ln \tilde{p}(A_{n_x,\hat{n}_t}=a)}{\hat{n}_{t} n_x}\hat{\tau}_{t}\tau_x
 \label{eq:strf}
\end{equation}

\begin{figure}[ht!]
  \centering
  \includegraphics[width=.8\textwidth]{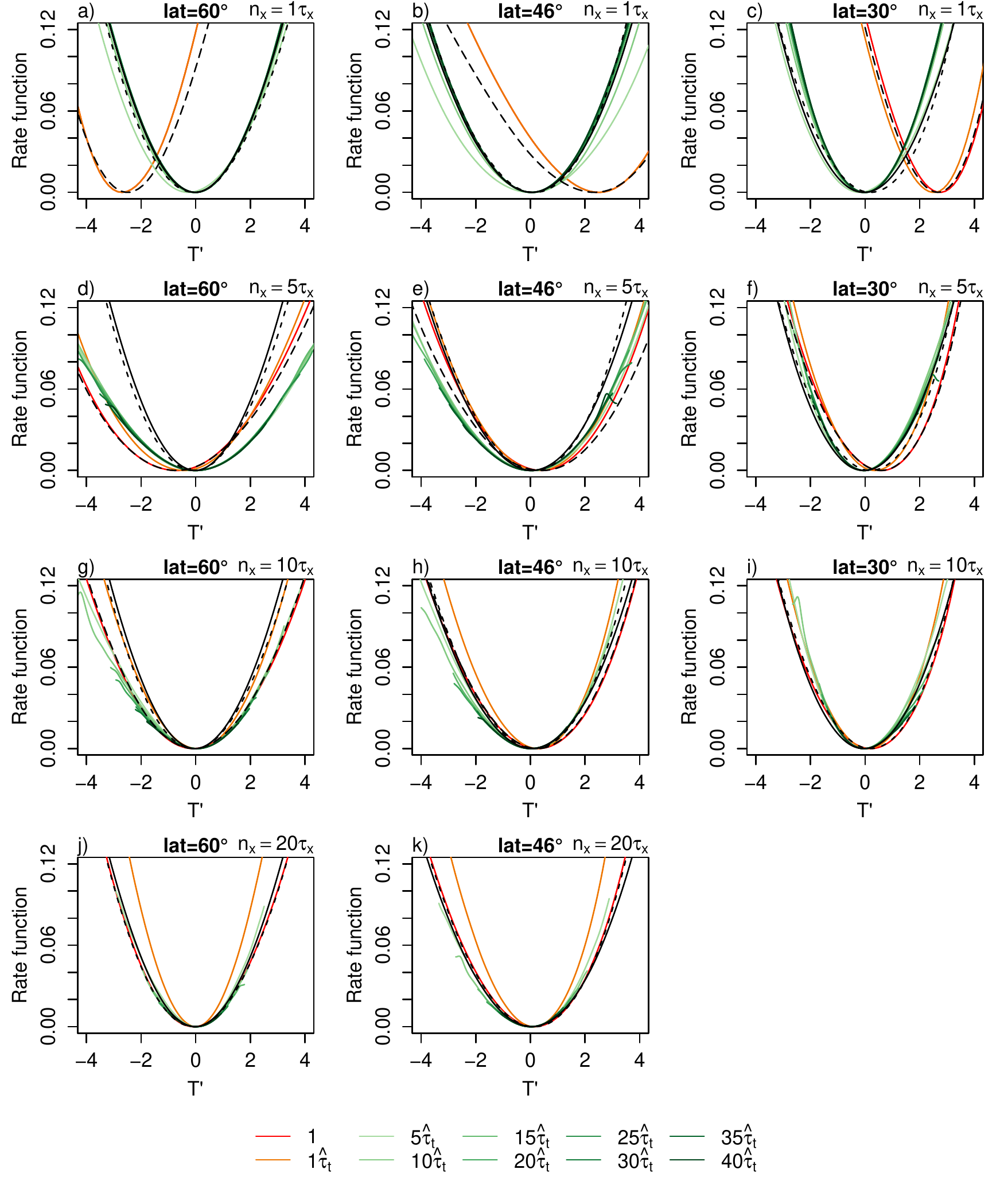}
  \caption{Rate functions of spatio-temporal averages for the selected latitudes and different zonal averaging lengths: a) -- c) $n_x=1\tau_x$, d) -- f) $n_x=5\tau_x$, g) -- i) $n_x=10\tau_x$, j) -- k) $n_x=20\tau_x$. The coloured lines represent spatio-temporal rate functions for different temporal averaging lengths $\hat{n}_{t}$ according to the legend. The black continuous line is the best temporal rate function estimate, the black short-dashed line is the best zonal rate function estimate, and the black long-dashed line is the zonal rate function estimate at the selected $n_x$. The rate function estimates are shifted vertically so that their minimum is at 0. $T'=T-\mu$ represents temperature fluctuations around the mean.} 
  \label{fig:strf}
\end{figure}

We remark that Eq.~(\ref{eq:strf}) accounts for both zonal and temporal auto-correlations by multiplication with $\hat{\tau}_{t}\tau_x$, similarly to the case of temporal and zonal rate functions. The spatio-temporal re-normalised rate function estimates are displayed by Fig.~\ref{fig:strf} (coloured lines). For comparison reasons, we also show the best temporal and zonal estimates $\tilde{I}_{n_t}^*$ (continuous black lines) and $\tilde{I}_{n_x}^*$ (short-dashed black lines), together with the estimate of the zonal re-normalised rate function at the selected zonal averaging length $\tilde{I}_{n_x}$ (long-dashed black lines). The main message here is that:
\begin{itemize}
\item The spatio-temporal re-normalised rate function seems to be equal to the universal function $I_n$ for \textit{small} and \textit{large} zonal averaging lengths. 
\item We suppose that in case of \textit{small} zonal averaging lengths $n_x\ll n_x^*$, like $n_x=1\tau_x$, the zonally averaged observable is not significantly different from the spatially localised observable, so that $\tilde{I}_{n_x,\hat{n}_{t}}$ converges to the universal function $I_n$. 
\item In case of \textit{large} zonal averaging lengths $n_x\ge n_x^*$, like $n_x=20\tau_x$, the zonal averages already exhibit universal characteristics, which are not altered by the additional temporal averaging, thus $\tilde{I}_{n_x,\hat{n}_{t}}$ corresponds again with the universal function $I_n$. 
\item On \textit{intermediate} levels however, i.e. $\tau_x < n_x \le n_x^*$, due to the non-trivial zonal correlations one obtains after zonal averaging a totally different observable from the original one. 
\end{itemize}

\begin{figure}[ht!]
  \centering
  \includegraphics[width=.5\textwidth]{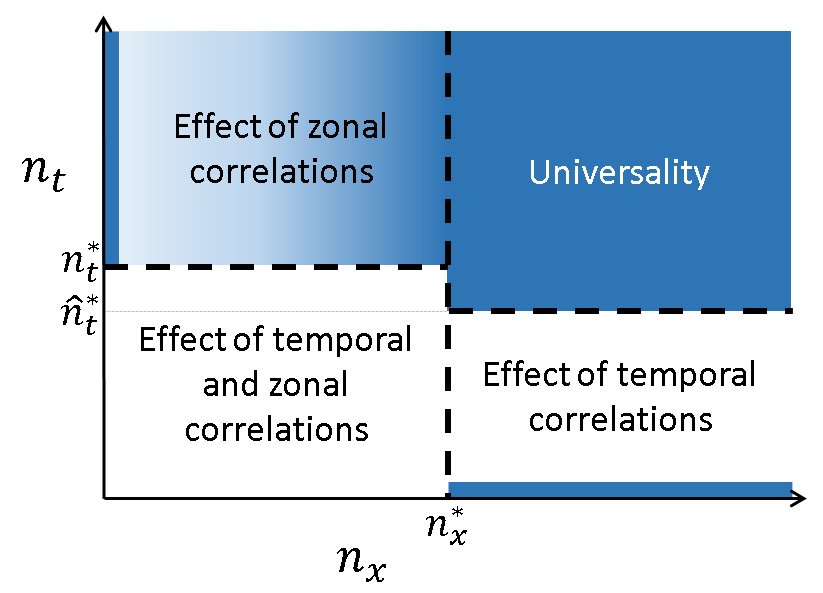}
  \caption{Schematic representation of universality and effect of correlations depending on the zonal and temporal averaging lengths. The dark blue colour marks the region where universality emerges. The light blue colour represents the region with non-universal spatio-temporal rate functions as an effect of zonal correlations. Pre-asymptotic regions, i.e. where the large deviation law is not valid yet, are white.} 
  \label{fig:scheme}
\end{figure}

The re-normalised spatio-temporal rate functions are different from the universal function $I_n$ at $n_x=5\tau_x$ in case the of latitudes $60\degree$ and $46\degree$, as well as at $n_x=10\tau_x$ in the case of latitude $60\degree$, whereas in this last case is worth mentioning that the spatio-temporal rate function corresponds with the zonal rate function estimate at $n_x=10\tau_x$. In all these cases, the spatio-temporal rate functions are flatter than the universal function, pointing out a higher probability of large deviations, which favours the presence of organised structures in the form of persistent weather patterns. Indeed, this is a new way to assess the existence of specific dynamical mechanisms - which result into distinct statistical properties of the temperature fields - associated to the low-frequency variability of the atmosphere discussed in the introduction.
Fig.~\ref{fig:scheme} represents schematically the ranges of temporal and zonal averaging lengths, at which universality emerges (blue) or is hindered (light blue) due to zonal correlations. Pre-asymptotic regions, where the large deviation law is not valid yet, are depicted as white.

As a side note, the horizontal shift of the rate function estimates at small averaging lengths ($n_x$ or $\hat{n}_t$) in Fig.~\ref{fig:strf} emphasises that these estimates are not reliable because the averaging length is too small for the law of large numbers to hold. We also wish to remark that differences emerge when looking at temperature data from latitude $30\degree$. Here, the spatio-temporal re-normalised rate function $\tilde{I}_{n_x,\hat{n}_{t}}$ at $n_x=1\tau_x$ is not identical to the universal function $I_n$. One possible reason for this is that when averaging over a length $n_x=1 \tau_x$ the newly defined observable has already significantly different properties from the local (in-space), time-dependent observable. The universality of the spatio-temporal rate function cannot be checked properly due to the limit in zonal averaging length of $10\tau_x$. What we see, however, is that at $n_x=10\tau_x$ spatio-temporal re-normalised rate function is quite similar to - yet distinct from - the universal function.

\subsection{Return Levels of the Large Deviations}
\label{subsec:return}

We summarise shortly our main findings presented until now:
\begin{enumerate}
 \item When considering temporal averages, the estimates of the rate functions seem to converge to an asymptotic function, and we obtain the best estimate of the rate function at an optimal averaging block length $n^*_t$. We show that there is a large deviation principle, i.e. a \textit{universal law} that allows us to estimate the probabilities of occurrence of averages over $n_t\ge n^*_t$, without having to actually perform the averaging.
 \item Spatial averages of the temperature field along latitudes obey the same large deviation law obtained for temporal averages. This means that we can deduce statistical properties of temporal averages from the ones of spatial averages and vice-versa. Additionally, the same asymptotic law is obtained when performing long spatial and temporal (two-dimensional) averages. 
 \item The temporal averages of temperature fields averaged on intermediate spatial scales along latitudes obey different large deviation laws, which point at a relatively higher probability of occurrence of heat waves and cold spells. This indicates the preferential existence of organised spatial structures, which correspond to the well-known low frequency variability of the atmosphere.
\end{enumerate}

\begin{figure}[ht!]
  \centering
  \includegraphics[width=.8\textwidth]{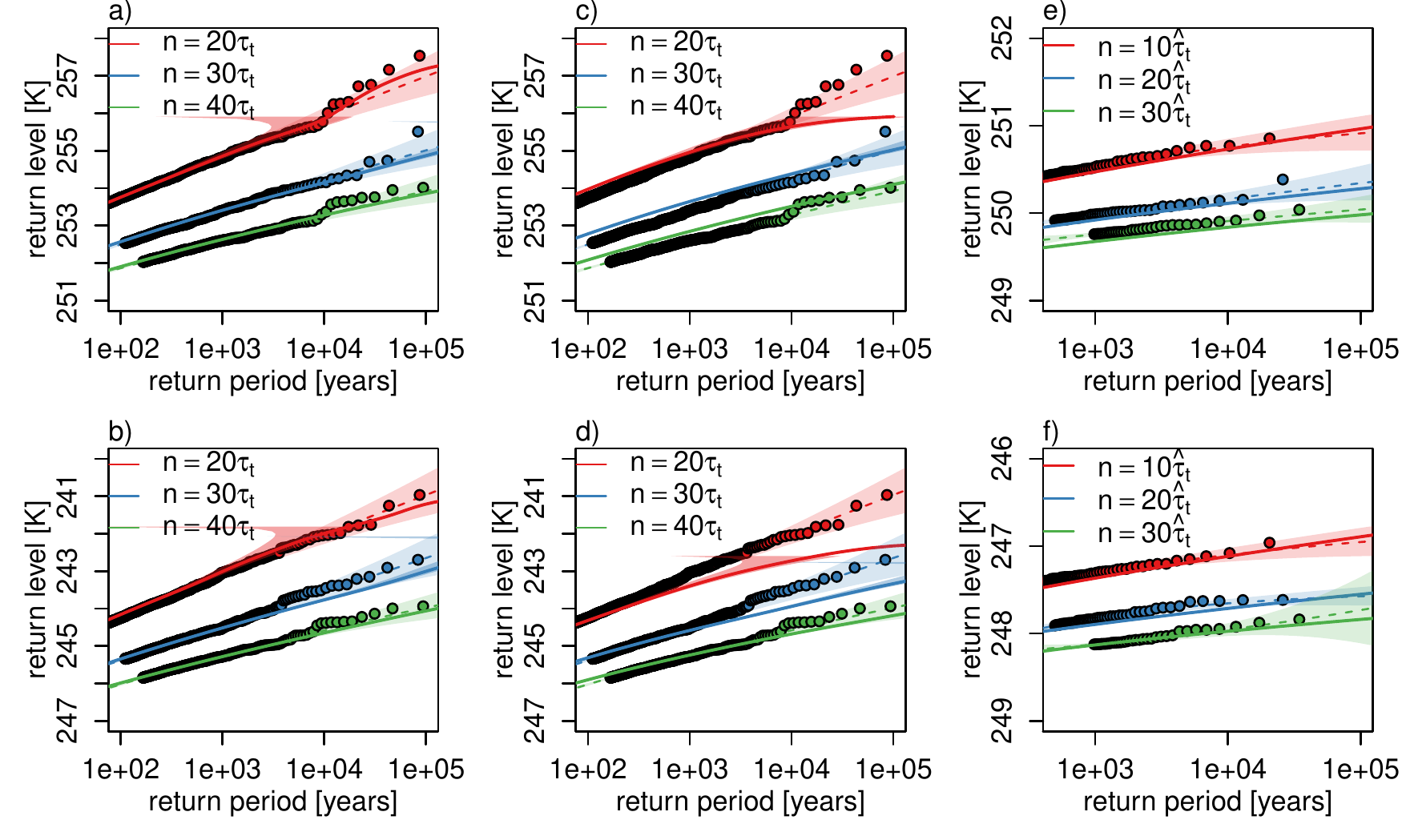}
  \caption{Return levels and return periods of positive (upper row) and negative (lower row), a) -- d) temporal and e) -- f) spatio-temporal large deviations of temperature at latitude $60\degree$. Circle markers: empirical data; continuous line with shading: estimates based on large deviations with 95\% confidence intervals of 2000 nonparametric bootstrap samples based on the normal distribution; dashed line with shading: Generalised Pareto estimates with 95\% confidence intervals based on Maximum Likelihood Estimation. The different colours represent different averaging lengths. The large deviation estimates are obtained based on a), b), e), f) temporal averages at $n_t^*=20\tau_t$, and c), d) zonal averages at $n_x^*=20\tau_x$.} 
  \label{fig:rl}
\end{figure}

Now, the question is how can we use these information in a practical way? One possible application, which we present in this subsection by the example of latitude $60\degree$, arises in the context of computing return periods of large events. Fig.~\ref{fig:rl} shows return level plots, i.e. return levels as function of return periods obtained in three different ways, based on: empirical data (circle markers), large deviation principle (continuous lines), and the Generalised Pareto distribution (dashed lines). For the estimation of return periods based on the large deviation limit, we first obtain kernel density estimates (function \texttt{density} of the \textsf{R} package \texttt{stats}, \citealp{r2016}) of the pdf's $p(A_{n}=a)$ at fixed equidistant return levels $A_{n,1},...,A_{n,256}$, based on which we estimate the cumulative distribution function $P(A_{n}\le a)$, and then compute the return periods for $A_n\ge a$ as $\frac{1}{1-P(A_{n}\le a)}$ and for $A_n\le a$ as $\frac{1}{P(A_{n}\le a)}$. Thus we obtain the return periods of both positive (Fig.~\ref{fig:rl}a,c,e) and negative (Fig.~\ref{fig:rl}b,d,f) large deviations.
The shading around the continuous lines in Fig.~\ref{fig:rl} represents the 95\% confidence intervals of 2000 nonparametric ordinary bootstrap return period estimates based on the normal distribution. 

We compute the Generalised Pareto return levels based on (\ref{eq:rlgpd}) using the maximum likelihood estimates of Generalised Pareto parameters (functions \texttt{gpd.fit} and \texttt{gpd.rl} of the \textsf{R} package \texttt{ismev}, \citealp{ismev2016}). We analyse return levels of high temperature values exceeding a threshold equal to the 99.9~\% quantile of the averaged series, as well as return levels of low temperature values below the 0.1~\% quantile. To verify the applicability of the peak over threshold method, the stability of return levels was checked also for a higher (lower) quantile of 99.99~\% (0.01~\%). The return level estimates seem to be stable even if the threshold is increased (not shown). Note that, although the very slow convergence of the Generalised Pareto shape parameter is well known in some cases, the stability of return level estimates still holds if the change in the shape parameter is relatively small as the threshold increases \citep{galfietal2017}. The shading around the dashed lines in Fig.~\ref{fig:rl} represents 95\% maximum likelihood confidence intervals of return level estimates. As a side note, in the case of the peak over threshold method the estimation concerns the return levels while the return periods are fixed, whereas we proceed the other way around in case of the large deviations. This is necessary because we estimate the rate function $I(a)$ at fixed equidistant values $a$.

In Fig.~\ref{fig:rl}a,b the return levels of temporal averages are shown for three different averaging windows $20\tau_t, 30\tau_t, 40\tau_t$. Here we use point 1. from above, and obtain the return periods based on the large deviation principle \textit{for every averaging window} from $p(A_{n_t^*=20\tau_t}=a)$. We notice a very good agreement with the empirical data and the Generalised Pareto return levels not only for $20 \tau_t$ but also in case of $30\tau_t$ and $40\tau_t$, for both high (Fig.~\ref{fig:rl}a) and low (Fig.~\ref{fig:rl}b) extremes of averages. In case of $n_t=20\tau_t$, the confidence intervals of the largest return periods based on large deviations become very unstable, the lower limits reach even negative values, thus they cannot be displayed on this semi-logarithmic scale. 

The return periods based on large deviations have an upper (or lower) limit because the estimation relies on empirical pdf's. This is not the case for the Generalised Pareto return periods since they can be extrapolated to even unobserved events. The large deviation principle, however, is a limit law that gives us return periods for every averaging length $n>n^*$, whereas the Generalised Pareto return periods have to be computed separately for every $n$. This becomes more and more difficult with increasing $n$ due to the decreasing data amount as effect of averaging. With other words, Fig.~\ref{fig:rl} points out the different dimensions in which the two limit laws act, as mentioned already in Sec.~\ref{sec:intro}. The predictability of the peak over threshold method (as well as generally of extreme value theory) is directed towards larger and larger events, i.e. towards unobserved ones, whereas the predictability of LDT is directed towards larger and larger averaging lengths, i.e. towards observables that, by construction, reduce dramatically the amount of data available for statistical analysis.

Point 2. presented above is illustrated by Fig.~\ref{fig:rl}c,d and Fig.~\ref{fig:rl}e,f. In the first case, return periods of temporal averages are computed based on the large deviation principle obtained for zonal averages ($n_x^*=20\tau_x$), and, in the second case, return periods of spatio-temporal averages (with a spatial averaging length of $20\tau_x$) are obtained from the large deviation law for temporal averages ($n_t^*=20\tau_t$). In both cases, but especially for the spatio-temporal averages, the agreement with the empirical data and the Generalised Pareto return levels is good. The differences between the return levels based on the large deviations and the empirical data (also Generalised Pareto return levels) are related to the discrepancies in the estimation of the temporal and zonal, as well as temporal and spatio-temporal re-normalised rate functions. For example, the underestimation of low extremes of temporal averages based on the zonal rate function has to do with higher re-normalised zonal rate function values compared to the temporal ones in their left tails (see Fig.~\ref{fig:rf1}g). 
We remark that the possibility of \textit{commuting} between averages of different dimensions (time and space) is due to the fact that by eliminating the effect of serial correlations the large deviations of these different dimensions follow a \textit{universal function}. 

\subsection{How sensitive are our results to the length of the numerical simulations?}
\begin{figure}[ht!]
  \centering
  \includegraphics[width=.8\textwidth]{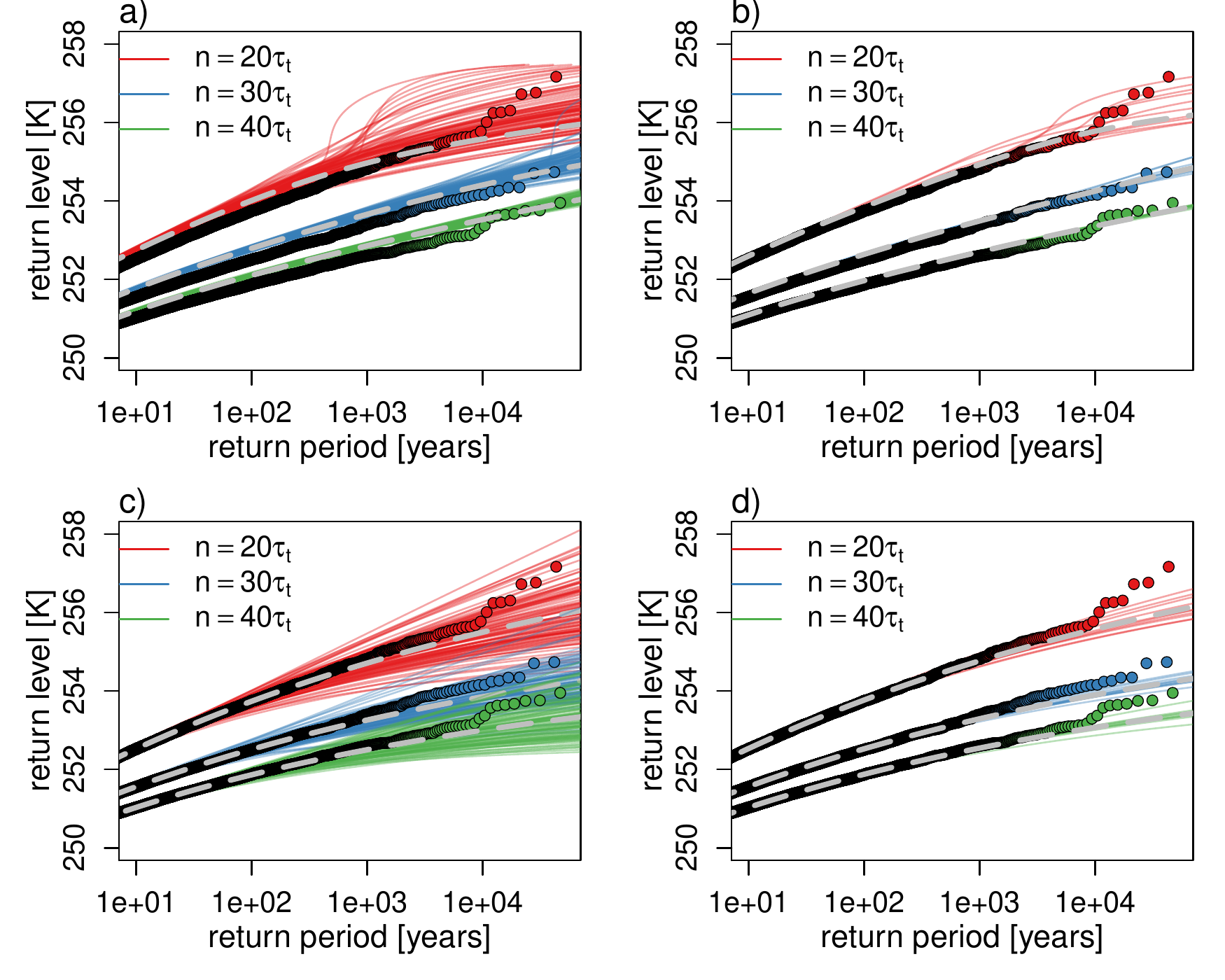}
  \caption{Return levels and return periods of positive temporal large deviations of temperature at latitude $60\degree$ based on temporal large deviation estimates at $n_t^*=20\tau_t$ obtained from a) 100 years and b) 1000 years simulations as well as based on Generalised Pareto parameters from c) 100 years and d) 1000 years simulations using the 95\% quantile as threshold. The different colours represent different averaging lengths. The circle markers represent the empirical data (10000 years) and the dashed grey lines illustrate estimate averages over the repetitions.}
  \label{fig:rl2}
\end{figure}

In typical data analysis exercises based on observational datasets or state-of-the-art climate simulations, the time span of available data is substantially less than in case of our idealized simulations, ranging from $O(100)$ to $O(1000)$ years. To test the applicability of the method in case of shorter time series, we divide our 10000 years long simulations into 100 pieces of 100 years simulations. For each of them we estimate return levels and periods of temporal averages based on temporal large deviations. We also estimate the Generalised Pareto return levels using the 95\% quantile as threshold. Afterwards, we increase the length of the simulations, and repeat these steps also for 10 pieces of 1000 years simulations. The obtained return levels and periods are illustrated by Fig.~\ref{fig:rl2}. Note that the empirical return levels are still obtained based on the whole amount of data of 10000 years, and thus represent a reliable basis of comparison.

Figure~\ref{fig:rl2}a demonstrates that climatologically important events with return periods of several tens and hundreds of years can be still approximated reasonably by large deviation estimates based on simulations of 100 years. The average of the estimates (grey dashed lines) slightly overestimates the empirical return levels, however the agreement is good until long return periods and improves with increasing averaging lengths. We additionally point out that the spread of predictions is remarkably low, and is decreasing with increasing averaging length. This underlines the advantage of using large deviation estimates for return levels of averages over large averaging windows, and shows that the predictions of LDT are very stable also if much shorter datasets are considered. The agreement with empirical return levels improves substantially by increasing the simulation length to 1000 years (Fig.~\ref{fig:rl2}b).

In case of the 100 years simulations, the Generalised Pareto return levels agree with the empirical data slightly better than the large deviation estimates (compare Fig.~\ref{fig:rl2}c with Fig.~\ref{fig:rl2}a). However, the variance of the Generalised Pareto estimates increases stronger with return period also for large averaging lengths. Furthermore, the averaged Generalised Pareto estimates (grey dashed lines in Fig.~\ref{fig:rl2}c) tend to underestimate the largest events for both 100 and 1000 years simulations. In case of realistic model simulations and observational data, however, one has to deal with additional complications besides the smaller amount of data, mainly as an effect of non-stationarity and strong correlations. We discuss these effects in the next section.

\section{Summary and Discussion}
\label{sec:disc}

We have analysed the properties of temporal and spatial near-surface (960 hPa) temperature averages in the PUMA simplified global atmospheric circulation model based on LDT. Extremes of averages on specific scales are related to persistent extreme events, like heat waves or cold spells. We run the model for 10000 years with a constant (only latitude dependent) forcing, creating non-equilibrium (due to the forced-dissipative nature of the model) steady state simulations without orography, annual or daily cycle. The forcing is symmetric for the two hemispheres. The horizontal resolution is $T42$ with 10 vertical levels, and the temperature values are recorded daily. We compute and compare the re-normalised rate functions based on the integrated auto-correlation for temporal and zonal temperature sequences at selected latitudes ($60 \degree$, $46 \degree$, and $30 \degree$), focusing on the mid-latitudes region, where turbulence is best developed. The spatial averaging is performed only in zonal direction, because this is the geometrical direction along which the system has a symmetry. We also analyse the case of two-dimensional, i.e. spatio-temporal averaging. We verify the correctness of our results by comparing the return periods based on the rate functions with return periods from the empirical data and from the peak over threshold method. Before discussing them in detail, we summarise first our main findings:

\begin{enumerate}
 \item The temperature averages in PUMA \textit{follow a large deviation principle}.
 \item The temporal and zonal re-normalised rate functions are equal if we compute them by eliminating the effect of temporal and zonal correlations. Thus, we can define a \textit{universal function}, describing temporal as well as spatial large deviations. 
 \item The spatio-temporal re-normalised rate functions are \textit{equal} to the universal function for \textit{small and large} spatial averaging lengths. \textit{On intermediate levels}, as an effect of non-trivial spatial correlation, the spatio-temporal re-normalised rate functions \textit{differ} from the universal one.
\end{enumerate}

Our results show that the temperature averages in PUMA follow a large deviation principle. The estimated rate functions clearly converge in case of temporal averages. We obtain reliable estimates at an optimal averaging length $n_t^*$, which is about $20 \tau_t$, where $\tau_t$ represents the temporal integrated auto-correlation. The fact that the temperature averages follow a large deviation principle might seem unsurprising, but actually it has extremely important consequences on a practical level. Based on large deviations, we can estimate the probabilities of averages, and thus for the practical use very important return periods, for every averaging length $n_t \ge n_t^*$. All we need to know is the probability of averages $A_{n_t^*}$, which we can estimate empirically. In contrast to the temporal averages, in case of zonal averaging the spatial averaging length $n_x$ is substantially limited by the size and shape of a latitudinal circle. The temporal averaging is performed on a theoretically infinite (and practically very long) line, whereas the zonal averaging takes place on a circle. Thus, the convergence of the estimated rate functions is not that clear as for temporal averages. However, the comparison of the zonal results with the temporal re-normalised rate function estimates shows that the averaging length $n_x^*=20 \tau_x$ seems to provide a reasonable rate function estimate, thus we choose this one as the optimal zonal averaging length. In case of latitude $30 \degree$, $20 \tau_x$ cannot be reached due to stronger zonal correlations. Here, the maximum averaging length is $10 \tau_x$.

We find that the temporal and spatial re-normalised rate functions seem to be equal if we eliminate the effect of correlations according to Eq. (\ref{eq:rfest}), where we basically scale the rate functions by the number of uncorrelated data instead of the whole number of data in an averaging block. Based on this equivalence, one finds a universal function $I_n=I_{n_t}=I_{n_x}$, in the sense that it describes both temporal and spatial large deviations. From a practical point of view, this implies that one can commute between space and time: we can deduce statistical properties of spatial averages (including return level estimates) from a single time series, and this is, of course, true the other way round too.

Obviously, based on a large deviation limit obtained in one dimension - time or space - we cannot describe persistent events, because the limit law is acting on very large scales, where spatial or temporal organisation is lost and universality emerges. However, as our results show, persistent space-time events can be studied based on LDT if one performs the averaging in both dimension - time and space.

Therefore, we extend our analysis also to spatio-temporal large deviations. Here, we average first in zonal direction taking different averaging lengths $n_x=1\tau_x,5\tau_x,10\tau_x,20\tau_x$, and then we search for a large deviation principle in time of the zonally averaged observables. We find that the spatio-temporal re-normalised rate function, computed again by eliminating the correlations according to (\ref{eq:strf}), is equal to the universal function $I_n$ in two cases: 1) for small zonal averaging lengths $n_x\approx \tau_x$, and 2) for large ones $n_x\ge n_x^*$. We suppose that in the first case, due to the small $n_x$, the zonally averaged observable is not significantly different from the temporal observable, and thus the rate function converges to the universal function. In the second case, the zonal averages already exhibit universal characteristics because the large $n_x$ allows for enough mixing in the series of zonal averages. These universal characteristics are not altered by the additional temporal averaging. On intermediate scales however, i.e. $\tau_x < n_x \le n_x^*$, due to the non-trivial zonal correlations, one obtains after zonal averaging a totally different observable, whose large deviations follow a clearly different rate function then the universal one. Consequently, by computing large deviations in time of zonal averages, we get rid of temporal persistence if the temporal averaging length is large enough, but we cannot eliminate the effect of zonal persistence on intermediate scales, which then leads to a non-universal re-normalised rate function. This also means that in this way we can study persistent extreme events based on LDT. These intermediate scales of about $5-10\ \tau_x$ or $\approx 2000 - 4000 \unit{km}$ are approximately equal with the scale of persistent synoptic disturbances, like the ones causing severe heat waves. According to this points of view, long lasting synoptic scale disturbances are large deviations from the steady state, which allow for a higher degree of spatio-temporal organisation and, in a loose sense, a lower entropy compared to disturbances at any other scales. This is an interesting signature of the so-called low-frequency variability of the atmosphere, which manifests itself in a complex phenomenology like in the case of blocking events \citep{tibaldimolteni2018}.

The advantage of applying LDT to analyse persistent climatic events is, besides the already discussed predictive power, the opportunity to learn something about the system under investigation:
\begin{itemize}
 \item Our system is chaotic enough to allow for a large deviation principle. This means that correlations decay sufficiently fast and the system is mixing enough for the chaotic hypothesis to hold. A very important characteristics of these kind of systems is that fluctuations are dominated by the mean instead of the biggest events, and thus the central limit theorem holds.
 \item The rate functions are approximately symmetric, so that positive fluctuations and correspoding negative flucutations of the same size have a similar probability of occurrence.
 \item We obtain an equivalence between temporal and spatial re-normalised rate functions, meaning that fluctuations in time are equivalent with fluctuations in space if one takes into consideration the different spatial and temporal scales. Thus our non-equilibrium steady state system exhibits a symmetry between the temporal and spatial (zonal) dimensions. This suggests that in the renormalized temporal and spatial dimensions the statistical properties of temperature can be considered as isotropic.
 \item We find that the spatio-temporal rate function related to intermediate spatial scales is substantially flatter and lower than the universal function. Consequently, large deviations in our system are more probable to appear on intermediate spatial scales than on any other scale.
\end{itemize}

Additionally, we compare the two frameworks for investigating rare events, i.e. LDT and the peak over threshold approach of extreme value theory, from a practical point of view, based on return level and return period estimates. Both methods are based on limit laws, but they differ in the way the limit is obtained, and thus also in the direction in which the limit acts. The peak over threshold approach deals with the conditional probabilities of averages exceeding a high threshold. The limit law is obtained as one considers larger and large extremes, thus it is directed towards large, even unobserved events. In case of LDT, we approach the limit as we consider averages with increasing averaging length $n$, thus the limit is directed towards $n \to \infty$. Our results point out these differences. On the one hand, the return level estimates based on the theory of large deviations are limited from above at small averaging lengths because they are obtained based on empirical distributions, whereas the estimates based on the peak over threshold approach can be extrapolated to unobserved events. On the other hand, the return levels based on large deviations can be obtained for every $n \ge n^*$ based on the probabilities of $A_{n^*}$, whereas in case of the peak over threshold approach they have to be estimated for every $n$ separately. We also have to remark that the convergence to the limit law seems to be easier to achieve in case of large deviations than in case of extreme values \citep{galfietal2017}. 

As mentioned above, we eliminate the effect of correlations in the computation of the rate functions by multiplication with the integrated auto-correlation. We estimate both temporal and zonal integrated auto-correlations, $\tau_t$ and $\tau_x$. By computing the ratio between spatial and temporal persistence, we define a scale velocity $U_\tau=\tau_x/\tau_t$, which is a measure for the anisotropy between space and time. If the anisotropy between space and time is strong, it becomes more difficult to show the existence of a universal rate function, as in the case of latitude $30 \degree$. We remark that the scale velocity we find by such asymptotic procedure could be viewed in connection with the research lines aiming at identifying the multifractal nature of the weather and climate fields \citep{lovejoyschertzer2013}, and, in particular, of precipitative fields \citep{deidda2000}. Generally, the connection between spatial and temporal scales is given by some characteristic velocity. In the multifractal analysis of spatio-temporal precipitation fields, the temporal dimension is usually rescaled by the advection velocity to fit the spatial ones, as explained by \citet{deidda2000}. If the rescaled temporal and the spatial dimensions are isotropic the overall advective velocity is sufficient to describe the relation between spatial and temporal properties of the precipitation field. In the case of spatio-temporal anisotropy, however, the advective velocity is scale dependent. In this work we are searching for the connection between time and space in terms of rate functions, and we find that this space-time connection is described very well by the ratio between the spatial and temporal integrated auto-correlation, which we indicate as $U_\tau$. As mentioned in Sec.\ref{sec:results}, $U_\tau$ is comparable with the zonal mean velocity at latitudes $60 \degree$ and $46 \degree$, which, indeed, advects turbulent structures at a first approximation. However the agreement is worse at $30\degree$. In this case, the dynamics has a mixed tropical/extratropical character, and long spatial correlations are due to the presence of the Hadley cell downdraft.

While nature and society do not typically conform to the hypotheses of the theorems needed to establish universal laws, such asymptotic results can nonetheless be extremely useful for studying observational data, just as in the widely case of extreme value theory. Therefore, this work should be seen as a first step towards the use of LDT for the analysis of actual climatic data and the outputs of state-of-the art climate models. The perspective is to find new ways to estimate efficiently the probability of occurrence of extremely rare events associated to persistent climatic conditions. In this work, we have focused on time scale which are long compared to those typical of the atmosphere, but one can adopt the same methods for studying persistent events of multiannual scales, where the oceanic variability is, instead, essential. This has, potentially, great relevance for addressing the problem of assessing human and environmental resilience to the low-frequency variability of the climate system. 

In case of applications to state-of-the-art model simulations or observational data, one has to deal with various complications, which are absent in our idealised model simulations. These have to do mainly with 
the presence of non-stationarity in the time series - as a result, e.g. of the seasonal cycle, and, on longer time scales, of climate change - as well as the presence of multiple time scales in the climate system, which may lead to slow decay of correlations for some variables. Note that, for example, ocean surface temperature decorrelates more slowly than the temperature over land surface, as a result of the larger heat capacity of the active surface layer in the ocean.
As discussed above in Sec. 2, strong correlations can inhibit the convergence to the limit law, at least when finite-size datasets are considered. Pragmatic approaches for dealing with time-dependent systems can be adapted from what done in the case of analyses based on extreme value theory. One can eliminate a long term trend, and then look at the detrended data using LDT. In a similar maner, it is possible to eliminate the annual cycle from the time series, obtain a large deviation principle, and consider the annual modulation later in the estimation of return levels. Another possibility would be to divide the time series according to seasons, and to obtain separate rate functions for separate seasons. Furthermore, it would be also more difficult to obtain universal properties of large deviations due to the high spatial heterogeneity as an effect of orography. However, we expect that this kind of universality should be found in regions with similar orographic and climatic characteristics.

Based on our idealised simulations, the estimated rate functions for the temperature fields are symmetric, suggesting that positive large deviations of temperature have the same probabilities as negative ones. In more realistic data sets however, we expect to find more frequently asymmetric rate functions. An argument supporting such a conjecture is that positive large deviations of air temperature should differ from negative ones in the presence of moist processes due to different chemical and physical characteristics of warm air compared to cold air, which has a much lower water vapour content. It might in fact be interesting to compare the large deviation rate functions of the surface temperature with those of the wet-bulb temperature, which takes into account the presence of moisture and is relevant for assessing heat stress \citep{zahidetal2017}. An alternative way to combine information on temperature and moisture is to look at the so-called equivalent potential temperature, which is proportional to the logarithm of the specific entropy of the air \citep{holton2004}. Another promising field for the application of LDT to geophysical data is related to precipitation induced landslides in mountain areas, where the standard modelling approach is the exceedance of a threshold defined by the cumulated rainfall intesity and duration \citep{keeferetal1987,peruccaccietal2017,ragnoetal2018}. We will leave the exploration of these research lines to future investigations.



\section*{Acknowledgements}

The authors wish to thank T. B\'odai for intellectual exchanges on the properties of extremes in geophysical flows, and F. Bouchet for stimulating conversations on the use of LDT for the study of geophysical flows. We would like to thank also E. Kirk and F. Lunkeit for providing help with the model simulations. VL and VMG acknowledge the support of the Sfb/Transregio project TRR181 "Energy Transfer in the Atmosphere and Ocean". VL acknowledges the support of the Horizon 2020 project "Blue Action". VMG acknowledges funding from the International Max Planck Research School on Earth System Modelling (IMPRS-ESM).



\newpage
\pagenumbering{Roman}						
\setcounter{page}{1}

\renewcommand\bibname{References}
\bibliography{references}


\end{document}